# Frontier Orbital Degeneracy: A new Concept for Tailoring the Magnetic State in Organic Semiconductor Adsorbates

**Anubhab Chakraborty,[1] Percy Zahl,[2] Qingqing Dai,[1] Hong Li,[1] Torsten Fritz,[1,3] Paul Simon,[3] Jean-Luc Brédas,[1] and Oliver L.A. Monti[1,4*]**

[1]Department of Chemistry and Biochemistry, University of Arizona, Tucson, Arizona 85721, United States

[2]Brookhaven National Laboratory, Center for Functional Nanomaterials, Upton, New York 11973, United States

[3]Friedrich Schiller University Jena, Institute of Solid State Physics, Helmholtzweg 5, 07743 Jena, Germany

[4]Department of Physics, University of Arizona, Tucson, Arizona 85721, United States

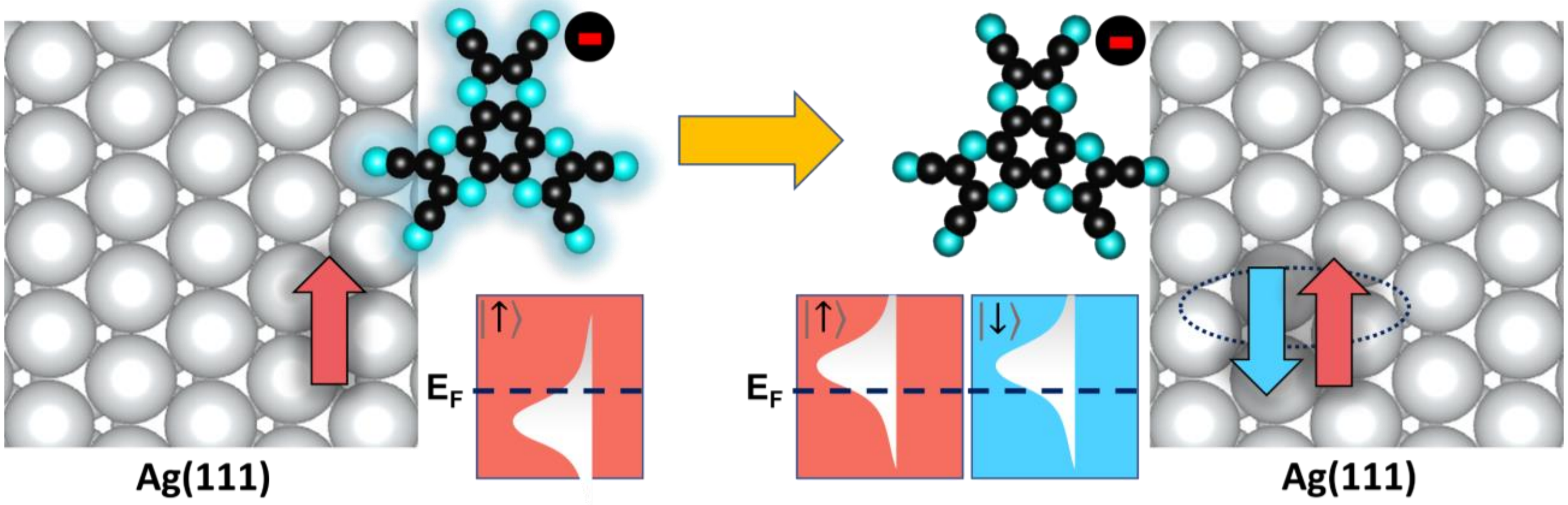

ABSTRACT: Kondo resonances in molecular adsorbates are an important building block for applications in the field of molecular spintronics. Here, we introduce the novel concept of using frontier orbital degeneracy for tailoring the magnetic state, which is demonstrated for the case of the organic semiconductor 1,4,5,8,9,11-Hexaazatriphenylenehexacarbonitrile (HATCN, $C_{18}N_{12}$) on Ag(111). Low-temperature scanning tunneling microscopy/spectroscopy (LT-STM/STS) measurements reveal the existence of two types of adsorbed HATCN molecules with distinctly different appearances and magnetic states, as evident from the presence or absence of an Abrikosov-Suhl-Kondo resonance. Our DFT results show that HATCN on Ag(111) supports two almost isoenergetic states, both with one excess electron transferred from the Ag surface, but with magnetic moments of either 0 or 0.65 $\mu_B$. Therefore, even though all molecules undergo charge transfer of one electron from the Ag substrate, they exist in two different molecular magnetic states that resemble a free doublet or an entangled spin state. We explain how the origin of this behavior lies in the twofold degeneracy of the lowest unoccupied molecular orbitals of gas phase HATCN, lifted upon adsorption and charge-transfer from Ag(111). Our combined STM and DFT study introduces a new pathway to tailoring the magnetic state of molecular adsorbates on surfaces, with significant potential for spintronics and quantum information science.

The highly versatile electronic structure of organic semiconductors has brought increasing attention in recent years to their potential application in the field of spintronics,[1–6] where the spin degree of freedom of electrons/holes is used in advanced opto-electronic and -spintronic devices and for new modalities of information processing. The diversity of readily accessible molecular scaffolds and ease of synthesis of derivatives allow precise control over the electronic structure, adsorption geometry, and the extent of wavefunction coupling of these molecules to different substrates.[7–9] This is crucial for tuning the electronic and magnetic interactions at the interface of molecular thin films on surfaces.[10] These underlying interactions are key to controlling the charge and spin states of molecules and offer an opportunity for designing molecular spintronic devices.

At an atomic or molecular scale, electronic correlations and many-body interactions are of great importance, particularly when considering the spin degree of freedom. An notable manifestation of such correlations is the Kondo effect, a many-body effect that arises from the coupling of a localized spin moment of an adsorbate to the electron spins in the conduction band of a metal substrate.[11–14] This gives rise to a singlet ground state due to screening of the localized spin moment by an itinerant cloud of spins from the Fermi sea of the substrate.[15] The signature of such Kondo screening is the appearance of a narrow resonance in the density of states (DOS) near the Fermi level, known as the Abrikosov-Suhl-Kondo (ASK) resonance.[14,16]

Original work on Kondo systems was concerned with transport anomalies in dilute magnetic alloys,[11] quantum dot systems,[17–20] and magnetic impurities,[13,21] however, in recent years, organic semiconductors adsorbed on coinage metal surfaces[22–27] have emerged as promising alternatives in this field because of their versatile chemical and electronic structures, consequently creating a rich parameter space for tailoring electronic correlations at interfaces. Multiple studies have explored the effects of chemical structure,[24,28] magnetic state of transition metal complexes,[27,29,30] adsorption geometry,[22,31] and self-assembly[32,33] in molecular Kondo systems. Particularly promising is the prospect of tailoring magnetic states in such systems, with initial attempts demonstrated in recent reports.[22,23,30,34,35] In all these works, transitions between the molecular doublet state and the molecule/substrate coupled singlet state have been achieved by manipulating the adsorption geometry or by modifying the chemical or electronic structure of the adsorbate molecule. The ability to *design* a molecular adsorbate system that is able to support both the molecular doublet state and the entangled Kondo singlet without chemical modification remains however an open problem.[1]

Here, we introduce a novel and likely general approach to meet this challenge: We show that an organic adsorbate whose frontier orbitals are degenerate can readily support multiple nearly isoenergetic magnetic states, and demonstrate that unidirectional Kondo switches can be realized in this manner. We use scanning tunneling microscopy and spectroscopy (STM and STS) of the organic semiconductor HATCN on Ag(111) together with density functional theory (DFT) to show that such magnetic bistability is intrinsic to this system due to the degeneracy of the lowest unoccupied molecular orbital of HATCN in the gas phase. Our work provides a novel concept to tailor magnetic states, and a new way to understand molecular Kondo systems. It unlocks new types of Kondo systems potentially suited for applications in information storage and computing by making use of tailored spin states in molecular adsorbates.

[1]*Department of Chemistry and Biochemistry, University of Arizona, Tucson, Arizona 85721, United States*

[2]*Brookhaven National Laboratory, Center for Functional Nanomaterials, Upton, New York 11973, United States*

[3]*Friedrich Schiller University Jena, Institute of Solid State Physics, Helmholtzweg 5, 07743 Jena, Germany*

[4]*Department of Physics, University of Arizona, Tucson, Arizona 85721, United States*

## Results and Discussion

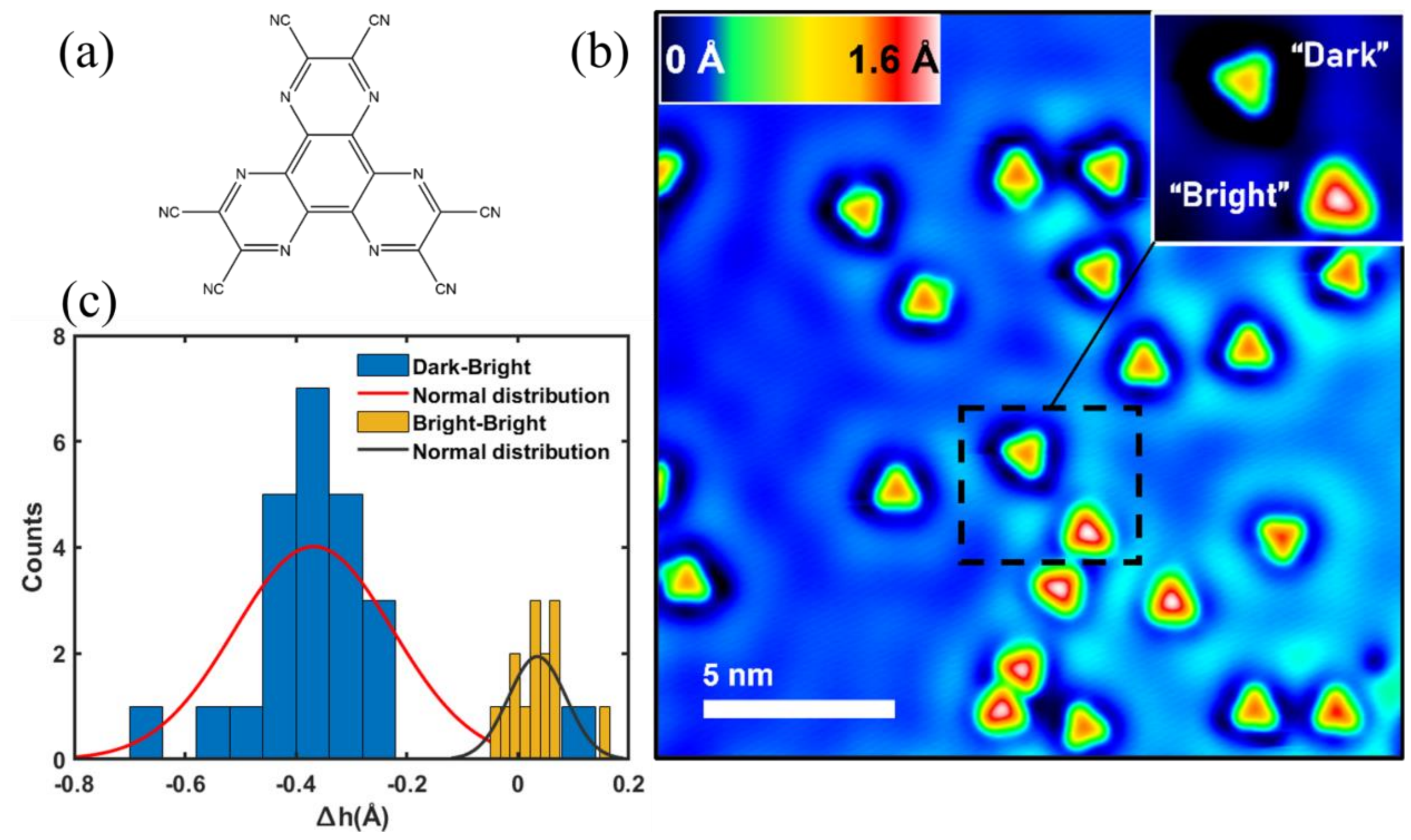

**Figure 1. (a) Molecular structure of HATCN. (b) Constant current STM image of HATCN molecules adsorbed on Ag(111) surface at 5 K (0.2 V, 0.01 nA). Inset shows a close-up of the two types of appearances. (c) Histograms showing STM height difference (Δh) between pairs of "Bright" and "Dark" types (blue) and pairs of "Bright" and "Bright" types (orange). Fits by normal distributions to the histograms are shown in red and black, and indicate an average Δ*h* of 0.37(2) Å and 0.03(1) Å, respectively.**

### *Observations*

We start by presenting a summary of the key findings obtained from scanning probe microscopy of HATCN on Ag(111). 0.1 monolayer equivalent (MLE, see Experimental Section) of HATCN molecules (molecular structure displayed in Figure 1a) adsorbed on an Ag (111) surface at 5 K show two distinct types of contrast in constant current STM images (Figure 1b): Molecules either have a bright center (labelled "Bright") or a less bright center (labelled "Dark", inset in Figure 1b) (see also Figure S1, Supplementary Information). Their appearance in this imaging mode is otherwise identical. Analysis of 50 pairs of HATCN molecules shows that on average the STM height of "Bright" molecules is 0.37(2) Å higher than that of "Dark" molecules (Figure 1c). Though we note that this height difference is an apparent height difference only, convoluted with the *z*-dependence of the tunneling matrix element,[36–38] we conclude from this and many other images that there are two and only two distinct molecular "types" of HATCN on Ag(111), distinguished by their apparent STM heights (see Figure S2, Supplementary Information).

To gain insight into the origin of the STM contrast differences, differential conductance ($dI/dV$) spectroscopy was performed. The $dI/dV$ spectra measured over a wide range of biases and acquired on the central benzene ring of HATCN (Figure 2a) show a distinctive narrow resonance/anti-resonance near the Fermi energy $E_F$ (i.e., bias of 0 V) in the case of "Dark" molecules (Figure 2b). This is distinctly different from the unstructured spectra near $E_F$ for the Ag(111) surface local DOS (LDOS) (Figure 2c). Interestingly, this zero-bias feature is exclusively observed in "Dark" molecules, and it is absent in "Bright" molecules (Figure 2a, b). Analysis of $dI/dV$ spectra for 60 molecules shows a 98% correlation between the molecule "type" and the appearance of a zero-bias feature in STS. As discussed in detail later, we interpret this narrow zero-bias resonance/anti-resonance, observed only on the "Dark" HATCN molecules, as a characteristic fingerprint of the Kondo effect, a result of screening in the HATCN/Ag(111) many-body system that manifests as an Abrikosov-Suhl-Kondo (ASK) resonance in transport measurements.[14,16,21,39–42] The requirement for the observation of an ASK resonance is the existence of unpaired electrons on the adsorbate, as is the case for HATCN: It accepts close to one full electron upon adsorption on Ag(111).[43,44] The line shape of the ASK resonance is described well by a Fano function, though other functional forms have been proposed as well.[45,46] This functional form considers the interfering pathways of (i) direct tunneling between tip and Ag(111) surface and (ii) tunneling mediated through the Kondo impurity (i.e., a HATCN molecule in the "Dark" state);[13,45,47,48] we use this model to fit the experimental $dI/dV$ spectra:

$$\frac{dI}{dV} = \frac{C(q+\varepsilon)^2}{1+\varepsilon^2} + \rho_0 \qquad (1)$$

where $\varepsilon = \frac{eV - E_0}{\Gamma}$, $E_0$ is the resonance energy, $2\Gamma$ is the full-width-at-half-maximum (FWHM) of the resonance, $q$ is the dimensionless coupling parameter between the magnetic impurity and the continuum states of the tip or the Ag substrate, and $\rho_0$ is the background differential conductance. Analysis of $dI/dV$ spectra for 35 HATCN molecules of "Dark" type gives $E_0$ and $\Gamma$ values of -7(1) meV and 38(1) meV, respectively. The values of $q$ range between ±1, likely due to different tip conditions between scans which lead to different tip-surface coupling, and indicating a significant role of the quantum interference. This is reflected in different peak shapes for the observed ASK resonance on the “Dark” HATCN molecules, as well as slight variations in the precise articulation of the HATCN appearance in the constant current STM images (see Figure S3, Supplementary Information).[48–50] The observation of different shapes for the zero-bias feature in different STS spectra (compare for example Figure 2b and Figure 3e) suggests that this is indeed an ASK resonance in the coupled HATCN/Ag(111) system and not a spectral feature of either HATCN or Ag(111). From the ASK resonance widths, the Kondo temperature is estimated to be 231 K.

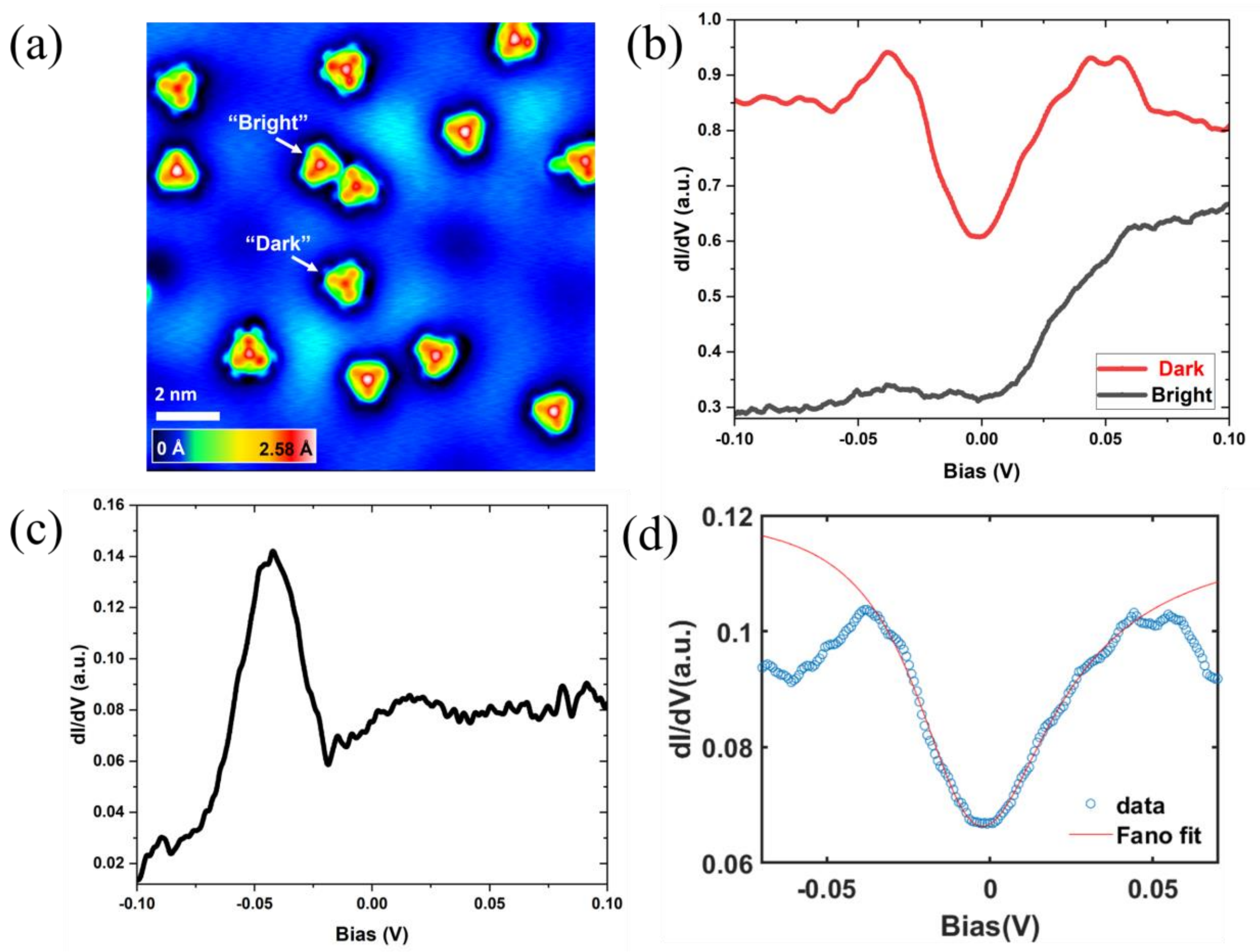

**Figure 2. (a) Constant current STM at 5 K (0.04 V, 2 pA) showing some of the molecules for which STS measurements were performed. (b) STS scan of molecules labelled in (a) demonstrating the existence of an anti-resonance feature for the “Dark” molecules and centered at -1 meV, in contrast to the unstructured spectrum for the “Bright” molecules. (c) $\frac{dI}{dV}$ spectrum of the Ag(111) substrate. (d) $\frac{dI}{dV}$ spectrum and Fano fit for the “Dark” molecule in (b), with $E_0$ = -1(1) meV, $\Gamma$ = 51(1) meV, and $q$ = -0.1(1).**

The small $E_0$ and narrow FWHM of the resonance are signatures of the ASK resonance and imply that “Dark” molecules act as spin-flip scattering centers embedded in the sea of Ag conduction electrons. Since the Kondo effect arises due to an interaction between an unpaired spin of an impurity with the physically neighboring electron sea,[11,22,49,51] we deduce that “Dark” molecules have an unpaired spin ($s$ = 1/2) while “Bright” molecules do not ($s$ = 0). This then points to the key feature of our system: *magnetic bistability* of HATCN on Ag(111), which we define as the coexistence of two nearly isoenergetic states with different magnetic configurations.

This is a remarkable finding for a molecular adsorbate system such as HATCN/Ag(111), since it is only correlated with the apparent height and not with the adsorption geometry, as we will also show later computationally. Previous studies have reported tunable control and activation of the Kondo effect in molecular adsorbate systems on metal substrates. This is usually achieved by using coordination chemistry between molecular units and paramagnetic atomic ions such as transition metals to create molecular alloys with intrinsic magnetic moments,[22,26–29,35,51–53] or by modifying the adsorption geometry of molecules using external stimuli such as voltage pulses,[23,24,26,31] temperature (via annealing),[22] or even molecular structure manipulation using an STM tip.[24,26] The present system is unique in this regard since the magnetic bistability is intrinsic to the system and does not require external stimulus or the presence of paramagnetic components.

Having established the existence of two magnetic states and consequently two distinct STM contrasts, we now demonstrate the ability of HATCN molecules to switch from a “Bright” to a “Dark” state. Figures 3(a) and (b) show an example of this effect: Observed in successive STM scans, we find that “Bright” molecules switch stochastically to the “Dark” state; this switching is unidirectional only, always going from “Bright” to “Dark”. It is important to mention that this is not a result of a significant displacement or rotation of the molecule, as evident from before and after STM scans and DFT calculations (*vide infra*). The switching is also not dependent on STM imaging conditions, which is confirmed by taking scans at different tip-sample biases and setpoint currents. Further, this is not a heat-induced effect either, as the sample was held at a constant 5 K throughout the measurements. The only change upon switching is of the molecular magnetic state, without affecting the adsorption geometry. To relate the change in appearance to the magnetic state of the molecule, we carried out differential conductance measurements on molecules that undergo switching. We observe that a switch from “Bright” to “Dark” (Figure

3c, d) is indeed accompanied by the appearance of a zero-bias Kondo resonance in the $dI/dV$ spectra (Figure 3e, red curve), as expected for a change in magnetic state. This implies that molecules can switch from a non-magnetic ($s$ = 0) to a magnetic state ($s$ = 1/2). Fitting the $dI/dV$ spectra for this particular molecule (Figure 3c, d) with the Fano model gives $E_0$ = -12(1) meV, $\Gamma$ = 38(1) meV, and $q$ = -0.2(1) (Figure 3f), consistent with the observed distribution of values for "Dark" molecules.

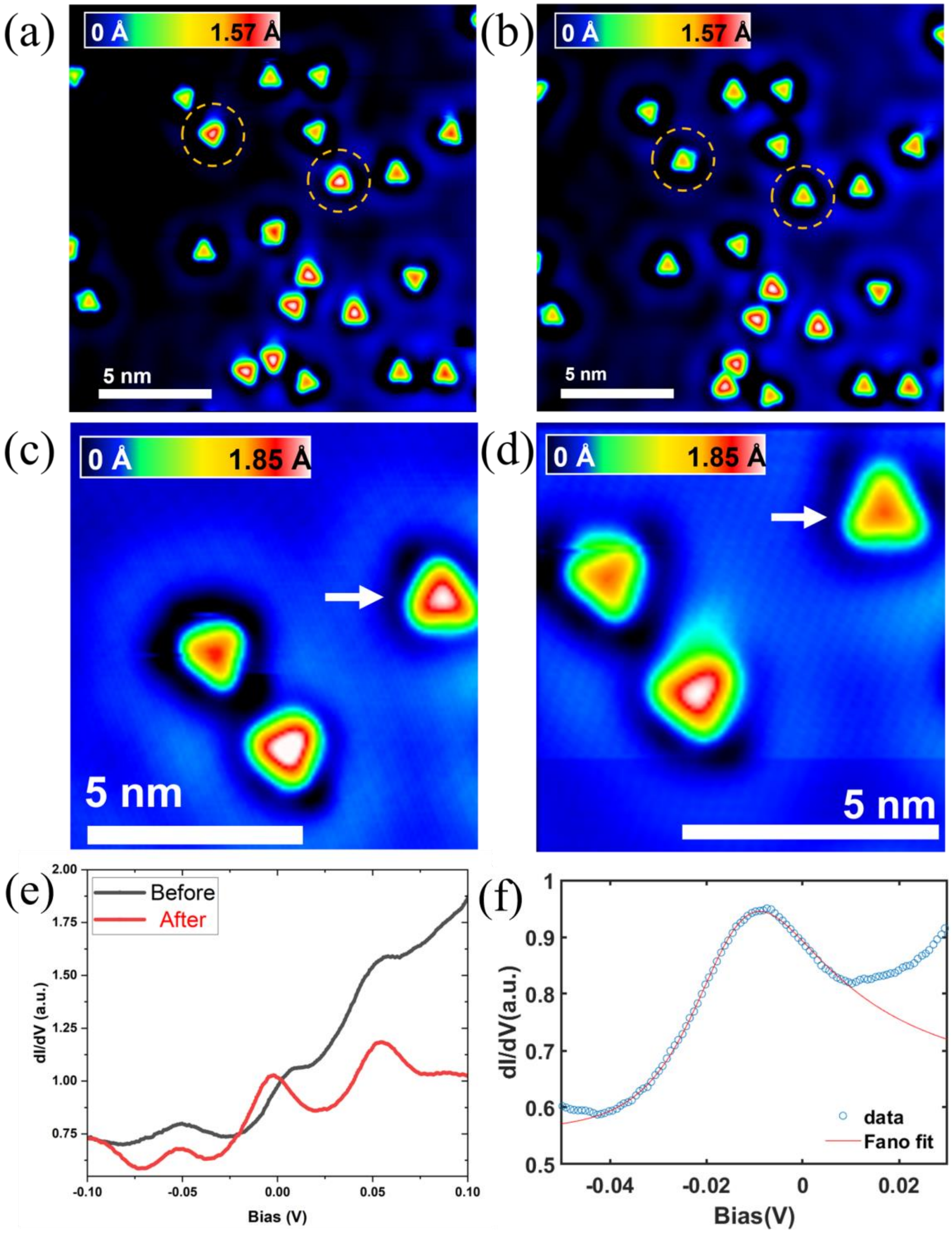

**Figure 3. Constant current STM images (0.2 V, 0.01 nA) (a), (c) before and (b), (d) after switching of highlighted molecules. (e) $\frac{dI}{dV}$ spectra of highlighted molecule in (c) and (d) before and after switching. (f) Fano fit for the "after" spectrum (red curve) shown in (e), with $E_0$ = -12(1) meV , $\Gamma$ = 38(1) meV, and $q$ = -0.2(1).**

*Origin of Magnetic Bistability*

To study whether the Kondo effect in the HATCN/Ag(111) system is limited to single molecules or can also be found in close-packed molecular layers, and to understand its origin, we also investigated the electronic structure and appearance by LT-STM and Scanning Tunneling Hydrogen Microscopy (STHM) in ordered molecular monolayers. As reported already,[43,44,54] HATCN forms well-ordered layers on Ag(111) that exhibit a honeycomb structure (Figure 4a) with 6 HATCN molecules arranged in a ring with a commensurate 7 x 7 alignment (see Figure S5, Supplementary Information).

For STS measured in the center of molecules in such a layer (same as for isolated HATCN molecules), we exclusively found spectra similar to those in Fig. 2b and Fig. 3e of the "Bright" state (black curves), i.e., without Kondo resonance (Figure 4b). We note that there is a very weak peak near 0 V for these molecules (Figure 4b), which does however not resemble the pronounced Fano-shaped features found in isolated "Dark" molecules. While the true nature of this small peak and another peak at +50 mV remains unknown, we speculate that these two features are related to the splitting of the doubly degenerate LUMO of isolated HATCN, with the lower level now partially occupied and the upper level sitting at +50 mV in both isolated and densely packed molecules. These features are caused by integer charge transfer from the Ag substrate irrespective of the molecular magnetic state (see computational results below). Also visible in these spectra is a peak around -50 mV which represents the surface state of Ag(111) (cf. STS curve of the bare Ag substrate in Fig. 2c; see also Figure S6, Supplementary Information).

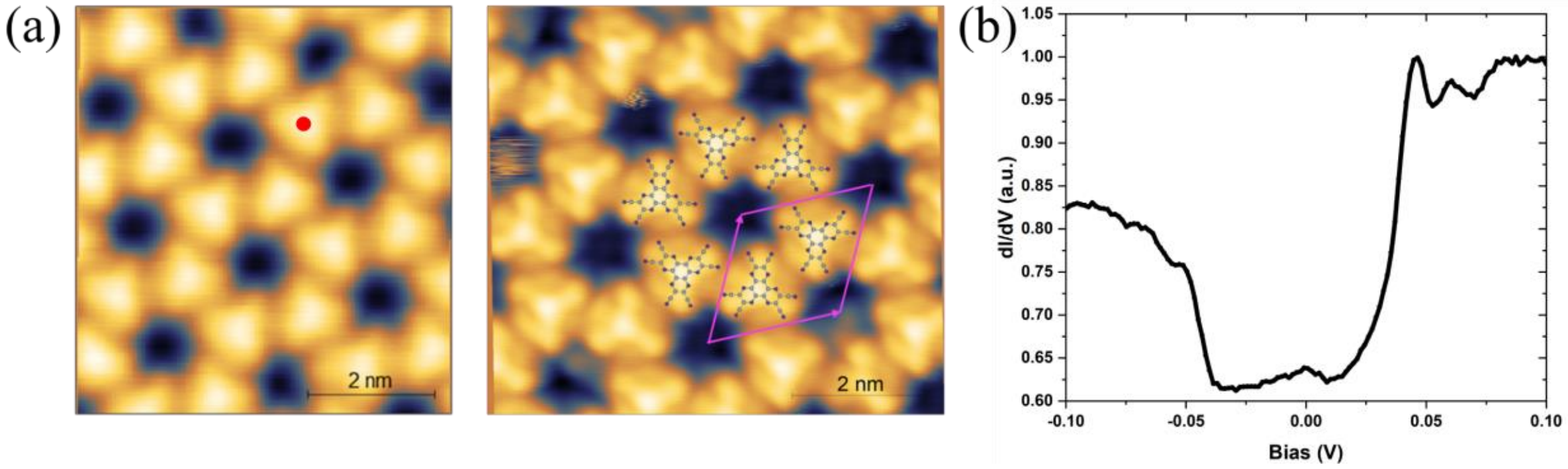

**Figure 4. (a) Constant current STM (left) (-0.4 V, 400 pA, 4.5 K) and STHM (right) (-0.5 V, 30 pA, 4.5 K) of a close-packed HATCN layer on Ag(111) showing the honeycomb structure. The colored red dot indicates the position where the $\frac{dI}{dV}$ spectrum shown in (b) was measured.**

We confirm therefore that the mere existence of a radical on a surface, while necessary, is not sufficient for observing a Kondo resonance. From those findings we conclude also that the Kondo effect is indeed limited to single molecules and does not occur in well-ordered honeycomb structures. Apparently, molecule-molecule interactions in the layers, either directly or substrate mediated, hinder the Kondo effect. This suggests that there exists a subtle balance between the two molecular states, “Dark” and “Bright” or Kondo and no Kondo. This is further indication of the near-degeneracy of the involved levels.

To understand the origin of this balance, we performed spin-polarized density functional theory (SDFT) calculations for the HATCN/Ag(111) system. Geometry optimizations indicate that the cyano groups are bent towards the substrate, causing a significant buckling of the molecule. Based on Hirshfeld charge analyses (when using the dDsC dispersion corrections, see Methodology) and Bader charge analyses (when using DFT-D3 dispersion corrections, see Methodology), the HATCN molecule accepts ~1 electron from Ag(111) upon chemisorption, irrespective of adsorption site or orientation on the surface. These findings are in line with previous works[44,54] and demonstrate that even “Bright” molecules are expected to have undergone charge transfer from Ag(111). Our SDFT calculations reveal the existence of two and only two different equilibrium spin configurations, in excellent agreement with our experimental observations of “Bright” and “Dark” molecules. We label these as “Charge Transfer (CT) Doublet” and “Charge Transfer (CT) Singlet”. The two configurations are very similar (see Table 1) in terms of their total energies ($\Delta E$ =10 meV) and the charge accepted from Ag(111) (~1 electron), which confirms that both types remain indeed radicals. We note that these results do not depend on the adsorption site. Remarkably, the “CT Doublet” configuration has a magnetic moment of 0.65 $\mu_B$, in contrast to the “CT Singlet” configuration (~0 $\mu_B$). This suggests that only the “CT Doublet” configuration, which is slightly more stable, is able to support an ASK resonance, while the “CT Singlet” configuration is overall non-magnetic despite its radical character. This interpretation is further supported by the spin-resolved projected density of states (PDOS), where the spin polarization of HATCN is maintained in the "CT Doublet" configuration with a split spin-resolved PDOS near $E_F$ (Figure 5a, red curves). Indeed, a spin-up singly occupied molecular orbital (SOMO) sits right at $E_F$, while the spin-down orbital is nearly completely unoccupied. In contrast, for the "CT Singlet" configuration the PDOS corresponding to the spin-up and spin-down states is nearly degenerate and straddling $E_F$ (Figure 5b, red curves). We note that the features near 0 eV in the PDOS plot for HATCN correspond to the HATCN SOMO and do not represent the ASK resonance observed in the $dI/dV$ spectra, since these DFT calculations cannot account for the many-body physics at play in the Kondo effect. However, the calculations demonstrate a salient feature of the magnetic bistability in this system: The spin associated with the excess electron accepted by HATCN from Ag(111) can be arranged in two different ways: (i) either "Up" ($|\uparrow\rangle$) or "Down" ($|\downarrow\rangle$), which gives rise to an ASK resonance and the "Dark" molecules, or (ii) in a superposition of

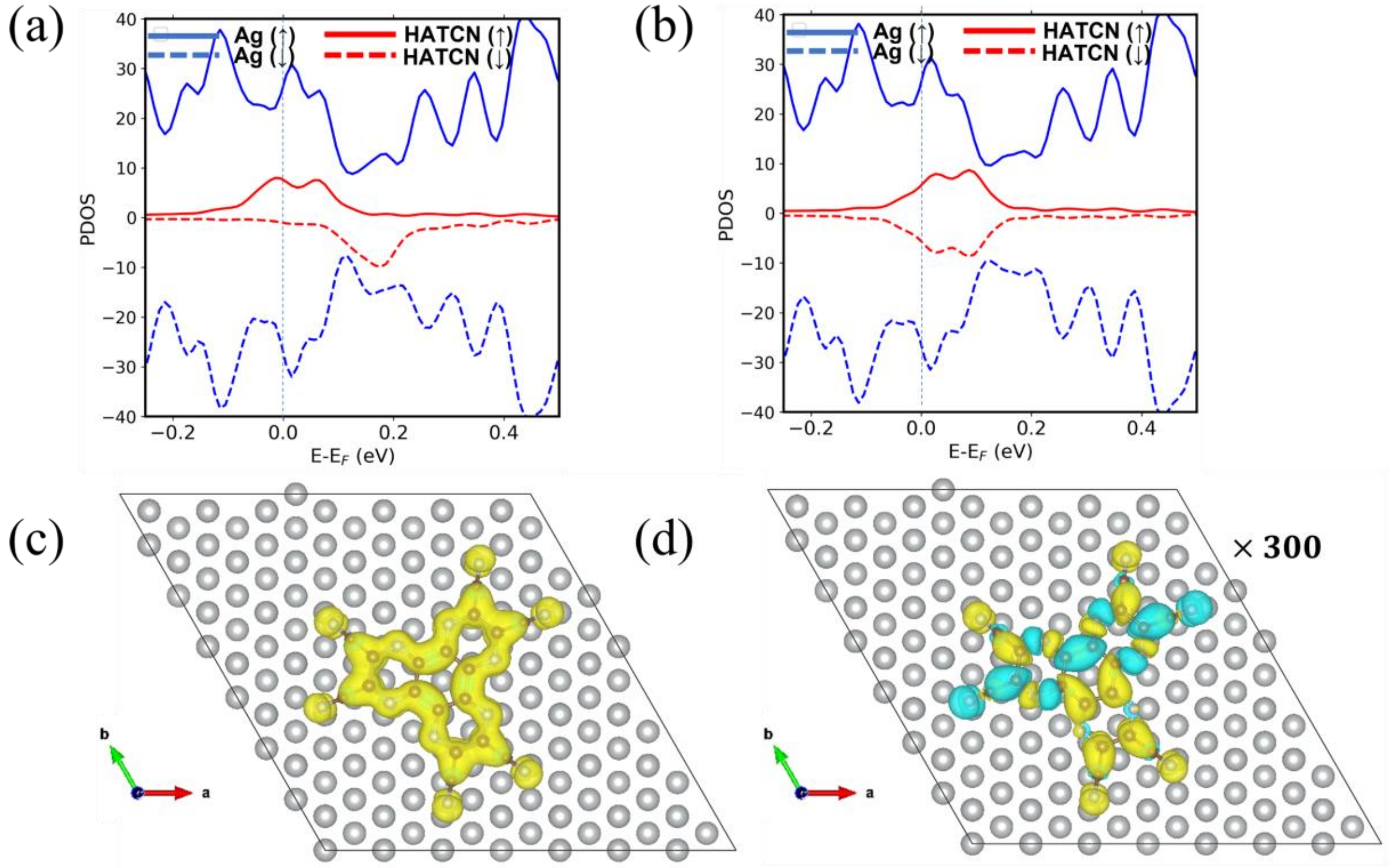

**Figure 5. PDOS of HATCN/Ag(111) for (a) CT Doublet and (b) ) CT Singlet configurations. Blue plots represent the Ag PDOS; red plots are HATCN PDOS. Spin-up (↑) and spin-down (↓) channels are shown using solid and dashed lines, respectively. Spin density plots for (c) CT Doublet and (d) CT Singlet configurations, with the isosurface value in (d) being 2 orders of magnitude smaller than in (c).**

“Up” and “Down” ($|\uparrow\rangle + |\downarrow\rangle$) with net magnetic moment equal to zero and without an ASK resonance, which is responsible for the “Bright” molecules.

**Table 1. Properties of the spin configurations in HATCN/Ag(111) from SDFT calculations**

| Configuration | Relative Energy (meV) | Magnetic Moment/$\mu_B$ | Charge Transfer (Hirshfeld) |
|---|---|---|---|
| **CT Doublet** | **0** | **0.65** | **0.98** |
| **CT Singlet** | **+10** | **~ 0** | **0.96** |

The associated spin densities for these configurations are plotted in Figures 5c and 5d, and demonstrate the presence of an electron spin delocalized over the whole molecule for the more stable “CT Doublet” version of the molecule, supportive of the observation of an ASK resonance. In contrast, the spin density for the “CT Singlet” molecule is nearly zero (note that the spin isosurface value is nearly 3 orders of magnitude smaller in Figure 5d than that in Figure 5c). As a consequence, even though this molecular adsorbate has also undergone electron transfer of nearly a full electron, spin entanglement leads to a net zero spin polarization and hence the absence of a prominent ASK resonance.

To understand this phenomenon in more detail, we analyzed the spin-resolved HATCN PDOS (Figure 5a, b) further to extract the energies of the molecular states near $E_F$ (Figure 6). For this analysis, we fit the HATCN PDOS with two Gaussians per spin channel, allowing the two-fold degeneracy of the LUMO in an isolated HATCN molecule to be lifted upon adsorption on Ag(111) and acceptance of an electron from the substrate (see Figure S7 and Section S10, Supplementary Information). From the area under the curve below $E_F$, we obtain the electronic charge for each spin. It is clear from this simple analysis that, even though the total electronic charge below $E_F$ is similar for both the CT Doublet (0.67 + 0.01 = 0.68 e) and the CT Singlet (0.50 e) configurations, the distribution of this total charge across different MOs is quite different in each case. As evident in Figure 6, for the CT Doublet, the charge resides in the spin Up channel ($|\alpha\rangle$) and in fact mostly in the $|\alpha_1\rangle$ state, leading to spin polarization. In contrast, the spin Up ($|\alpha\rangle$) and the spin Down ($|\beta\rangle$) channels remain degenerate for the CT Singlet configuration, and the total charge is distributed equally and almost entirely in the $|\alpha_1\rangle$ and $|\beta_1\rangle$ states. This leads to spin entanglement and net zero spin polarization. Note that these extracted charges are in reasonable agreement with the Hirshfeld charges, particularly considering that they are obtained from a simple model of the PDOS near $E_F$. This energy level diagram is also consistent with the experimentally obtained energies of HATCN/Ag(111) states. For instance, we can assign the features at +11 mV and +60 mV for the spectrum of the molecule "Before Switching" (or "Bright") (Figure 3e, black curve) as the CT Singlet $|\alpha_1\rangle$ (or $|\beta_1\rangle$) and the CT Singlet $|\alpha_2\rangle$ (or $|\beta_2\rangle$) states, respectively. For the spectrum of the molecule "After Switching" (or "Dark") (Figure 3e, red curve), the peaks at -12 mV and +54 mV are assigned to the ASK resonance and the CT Doublet $|\alpha_2\rangle$ states, respectively. We emphasize once more that the features near 0 V for "Bright" HATCN molecules (Figure 3e) and molecules in close-packed layers (Figure 4b) are assigned to the CT Singlet $|\alpha_1\rangle$ (or $|\beta_1\rangle$) state and are not the ASK resonance observed in the $dI/dV$ spectra of "Dark" HATCN molecules (see Section S6, Supplementary Information).

To summarize, the key ingredient for the observed magnetic bistability stems from lifting of the degeneracy of the LUMO upon adsorption of HATCN on Ag(111) and charge transfer to the molecule. Spin-spin correlation interactions with the surface lead to small but crucial energetic differences and orbital occupancies which manifest in fundamentally different magnetic states. Taken together, the STS and DFT results show that because of gas phase degeneracy of molecular frontier orbitals there exist two energetically close-lying states with fundamentally different magnetic properties for isolated HATCN molecules on Ag(111). It is conceivable that the rather small energetic difference of 10 meV is compensated by intermolecular interactions in HATCN films, hindering the observation of a Kondo resonance in that case.

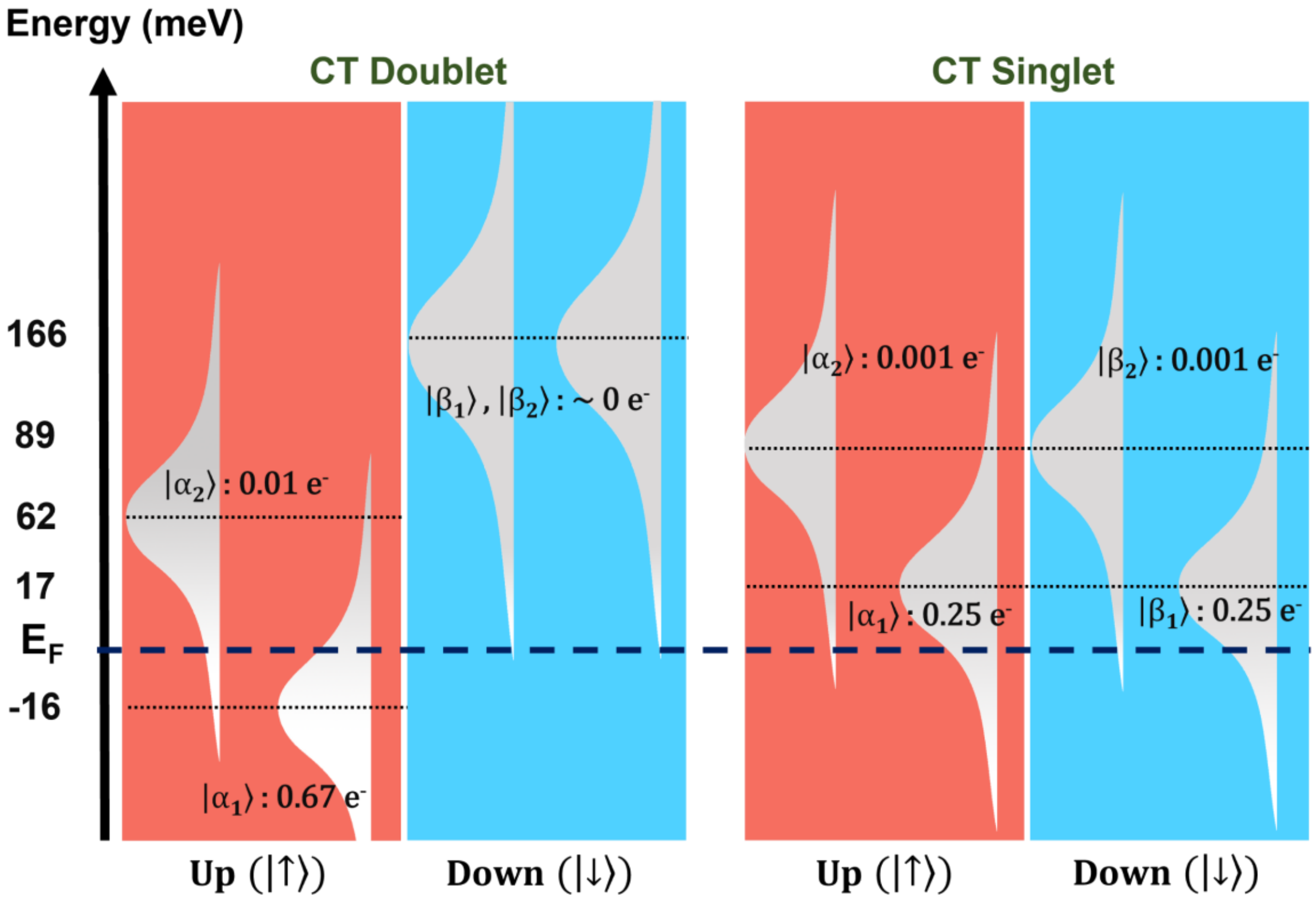

**Figure 6. Energy level diagram for the CT Doublet and CT Singlet configurations of 0.1 ML HATCN/Ag(111) near $E_F$, obtained from fitting the calculated HATCN PDOS (red curves in Figure 5a, b). The indices 1 and 2 refer to the two non-degenerate former LUMOs of adsorbed HATCN, while α and β represent the spin Up and the spin Down channels, respectively. Red and Blue boxes show the single electron states for Spin Up and Spin Down channels, respectively. In addition, the electron charge below $E_F$ is shown for each state.**

We also note that the process of acquiring an STM image itself implies that given the calculated small energy difference between the two configurations, switching by the STM tip, e.g., due to high electric fields in the tunnel junction, or spontaneous thermally activated switching can both be expected, consistent with our observations.

To investigate the validity of these conclusions, we considered a number of alternative hypotheses. First, we considered different adsorption geometries leading to different electronic configurations as a possible explanation for our experimental observations. We performed high-resolution atomic force microscopy (HR-AFM) imaging on the HATCN/Ag(111) system. Notably, we do not find any correlation between the HR-AFM and STM contrasts (Fig 7a, b), indicating that whatever geometric differences may exist, including adsorption sites or a small rotation on the surface, these do not explain the difference between "Dark" and "Bright" molecules. In fact, there is no consistent contrast difference between different HATCN molecules in the HR-AFM images themselves (Fig 7a and Figure S11, Supplementary Information) nor a correlation with the contrast differences seen in the STM scan (Fig 7b) to suggest the existence of two different HATCN adsorption geometries. HR-AFM provides information about chemical bonds and on-surface molecular structure,[26,55,56] and the lack of distinct differences correlated between STM and HR-AFM suggests that the origin of the two types cannot be based on different on-surface molecular structures. This is also in agreement with the fact that from DFT geometry optimizations obtained at different adsorption sites of Ag(111) (Fig 7c), there is no strict molecule-surface registry, and many translationally and rotationally slightly different configurations exist (see Figure S4, Supplementary Information). Importantly, these slightly different adsorption configurations all result in the same electronic structures and very similar magnetic moments (see Section S8 and S9, Supplementary Information), unable to explain the origin of the differences between "Bright" and "Dark" molecules observed in LT-STM and $dI/dV$ spectra. We also note that the higher spatial resolution of HR-AFM helps to exclude adatoms or atoms sandwiched between the Ag(111) surface and HATCN as potential causes for the observed contrasts and different magnetic properties.

Second, we investigated the possibility of two adsorption height minima in the adsorption energy. A hypothetical additional minimum at a slightly larger height could trap molecules in a metastable site, likely with different electronic properties and molecule-surface coupling due to the expected weaker electronic coupling to the substrate. Under the influence of an electric field or over time one would expect that such molecules escape this first minimum to reach the thermodynamic minimum on the surface. Our DFT calculations (Fig 7d) unambiguously show, however, that there are no other minima in the adsorption energy profile;[31,34] the two STM contrasts are therefore not caused by metastable adsorption geometries. Instead, the different STM apparent heights are primarily reflective of two different electronic structures and indeed magnetic states.

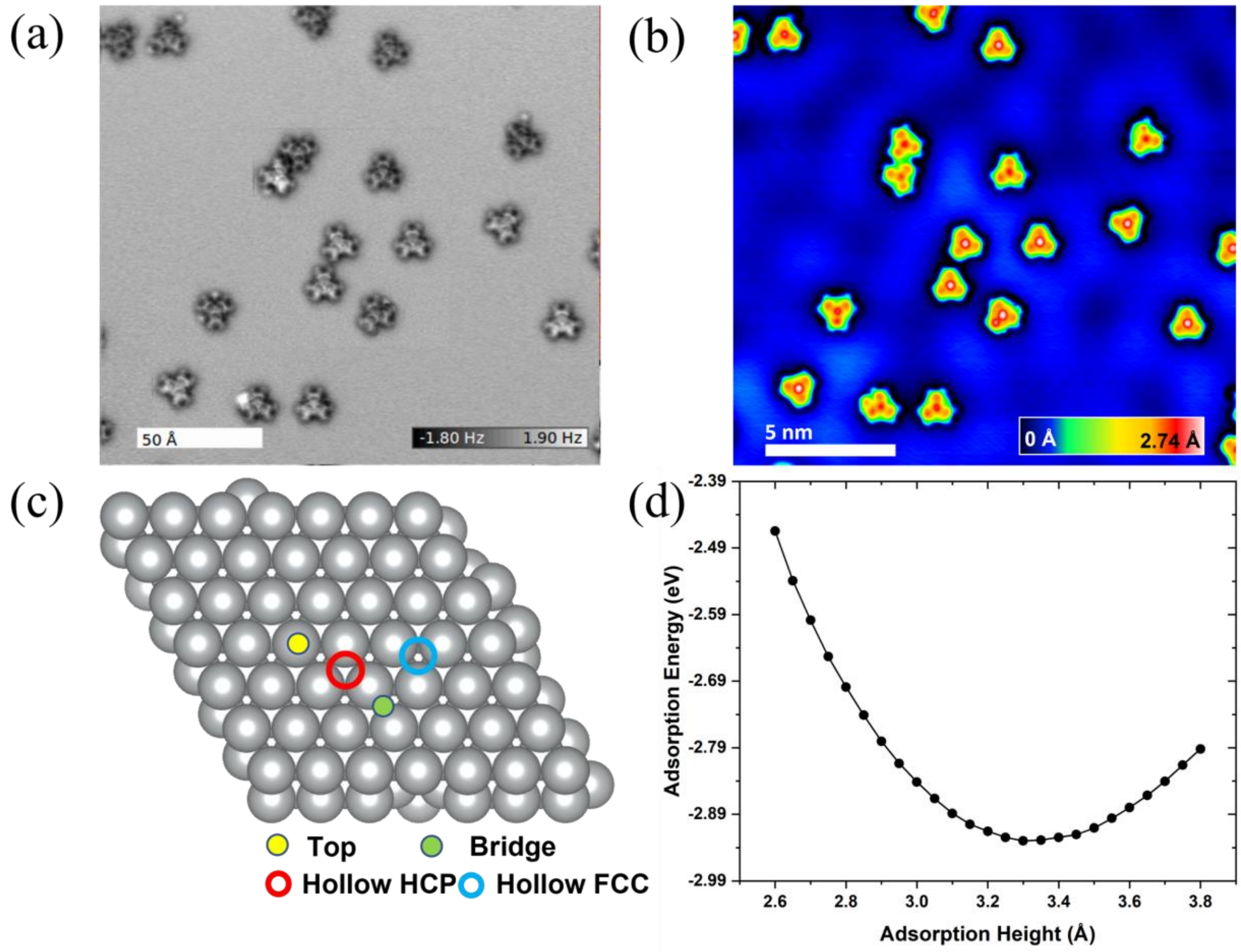

**Figure 7. (a) Constant height HR-AFM (0.02 V) and (b) constant current STM at 5K (0.05 V, 2 pA) images taken over the same region. The contrast seen in (b) showing the two types cannot be observed in (a). (c) Adsorption sites on Ag(111) considered in our DFT calculations. (d) DFT calculated adsorption energy profile for the Hollow FCC site. The minimum at 3.3 Å corresponds to the adsorption height of HATCN on Ag(111).**

Taken together, our DFT calculations indicate that the origin of the distinct STM contrasts cannot be explained by differences in HATCN adsorption on Ag(111). The combined results from DFT and HR-AFM strongly suggest that adsorption induced structural or electronic differences are minimal for the HATCN/Ag(111) system.

*Conclusions*

We close by highlighting the significance of our findings. The Kondo effect in molecular adsorbate systems has conventionally been manipulated by altering either the electronic structure or the adsorption geometry of the adsorbate species. The novelty of the HATCN/Ag(111) system presented here is that neither of these methods are required to support two different magnetic states on the surface. Instead, we propose that this kind of magnetic bistability and switching results from lifting of the frontier orbital degeneracy upon adsorption, and will be manifested in molecular systems with a minimum of two-fold degeneracy near $E_F$. In fact, a bi-directional switching might be possible at room temperature because of the energetic closeness of the CT Doublet and the CT singlet states as well as the estimated high Kondo temperature.

More broadly, the ability of the HATCN anion radical to exhibit two completely different spin distributions with only a small energy difference is quite intriguing and may allow for the creation of a Kondo lattice by preferentially "pushing" the molecule to one or the other magnetic state. Lastly, our results provide deeper insight into the Kondo effect by considering the spin density distribution and associated magnetic moments for all supported configurations in the radical species instead of only connecting the effect to the extra electronic charge. Our work therefore opens avenues for precise control of molecular Kondo switches, creation of Kondo lattices and tailoring of surface-adsorbate interactions in organic semiconductor interfaces.

*Experimental Section*

**Sample Preparation**

Ag(111) was cleaned using $Ar^+$ sputtering (1 KeV, 5 μA/cm$^2$) and annealing (823 K). Commercially acquired HATCN (Alfa Chemistry, 99.5%) was purified by three cycles of gradient sublimation (573 K) in a custom-built furnace. The UHV system base pressure was 7 x $10^{-10}$ Torr. HATCN molecules were directly deposited on the cold (5 K) Ag surface by flashing a small amount from a silicone carrier substrate that was heated by direct current. Film thickness is reported in terms of monolayer equivalent (MLE) as a fraction of a hypothetical monolayer (ML) on the Ag(111) substrate held at 5 K, with 1 ML ≈ $5.6 \times 10^{13}$ molecules/cm$^2$.[54] The last preparation step was exposing the sample to a small CO dose while at 5 K, specifically 10 s CO exposure at $2\times10^{-8}$ mbar as required for tip functionalization purpose only.

## STM-STS-AFM

### *0.1 MLE HATCN/Ag(111)*

Measurements were performed with a Createc-based low-temperature (LT)-STM system custom upgraded with HR-AFM capability and operated using the open source GXSM control software.[57–59] HR-AFM measurements were performed using GXSM's special constant height control mode with automated constant current (STM mode) transitions if a compliance setting (probe safety or also automated big/3D molecule lift mode) of a maximum allowed tunnel current is exceeded. A small bias of 20 mV was typically applied in HR-AFM-mode. For frequency detection the custom, high-speed GXSM RedPitaya-PAC-PLL controller was used.

STS was performed using an external Lock-In Amplifier (SRS Model 7265 Dual Phase DSP Lock-in Amplifier). The bias was modulated at 299 Hz at typically 10mV or 5mV pure sine amplitude. Other experimental parameters such as imaging conditions, stabilization point (bias voltage and tunneling current), and sample temperature are specified where necessary.

### *1 ML HATCN/Ag(111)*

STM and STHM measurements of HATCN layers on Ag(111) were performed with a JT-STM/AFM system from SPECS Surface Nano Analysis by using a mechanically cut PtIr tip. The STM system is operated with a Nanonis SPM controller (Version 5) featuring an integrated lock-in amplifier that we used for the STS experiments. Here, a typical modulation voltage of 6 mV with a tip lift of 1 nm was set. The working temperature was 4.5 K. The STS spectrum shown in Fig. 4b was taken as an average of several measurements from one STS map of a HATCN layer. Triplicate measurements were made at 11 molecular centers on different HATCN molecules and averaged. The averaging was done independently. Additional experimental parameters are given where necessary.

## DFT

Spin-polarized density functional theory (SDFT) calculations were carried out using the projector augmented wave (PAW) method[60] as implemented in VASP.[61,62] The Ag(111) surface was modeled by five-layer slabs with the bottom three layers fixed at the bulk Ag structures and the top two layers relaxed; the HATCN/Ag interfaces were modeled by a supercell consisting of 5x5 Ag(111) primitive cells and by considering adsorption sites of a HATCN molecule on hollow-fcc, hollow-hcp, top, and bridge sites. In the initial screening of all adsorption sites, the Perdew−Burke−Ernzerhof (PBE) functional was used with DFT-D3 [63]dispersion corrections; once the most stable adsorption site (hollow FCC) was found, the PBE functional with density-dependent energy corrections (dDsC) [64,65]was applied in further geometry optimizations and subsequent electronic-structure calculations. In all calculations, a cutoff energy of 500 eV was considered for the planewave basis set. A Γ-point only k-mesh and a 5x5x1 Monkhorst-Pack k-mesh were used in the geometry optimizations and electronic-structure calculations, respectively. In the structure optimizations, the atomic positions were relaxed until the forces on each atom were smaller than 0.01 eV/Å and the energies converged to $10^{-6}$ eV. A Bader (for DFT-D3 corrections) or Hirshfeld (for dDsC corrections) analysis was adopted to evaluate the charge transfer from the Ag(111) surface to the HATCN. Simulations of constant-current STM images were carried out in the framework of the Tersoff-Hamann approximation.[66,67]

## Author Contributions

Anubhab Chakraborty: conceptualization, data curation, formal analysis, investigation, methodology, and writing – original draft. Percy Zahl: data curation, investigation, methodology, resources, software, funding acquisition. Qingqing Dai: data curation, investigation. Hong Li: data curation, investigation, methodology, resources. Torsten Fritz: funding acquisition, project administration, resources, supervision, validation, and writing – review & editing. Paul Simon: data curation, investigation. Jean-Luc Brédas: funding acquisition, project administration, resources, supervision, validation. Oliver L.A. Monti: conceptualization, funding acquisition, project administration, resources, supervision, validation, and writing – review & editing.

## Corresponding Authors

**Oliver L.A. Monti:** Department of Chemistry and Biochemistry, University of Arizona, Tucson, Arizona 85721, United States; Department of Physics, University of Arizona, Tucson, Arizona 85721, United States; Email: monti@arizona.edu; Phone: (+1) 520 626 1177; orcid.org/0000-0002-0974-7253.

**Authors**

**Anubhab Chakraborty:** Department of Chemistry and Biochemistry, University of Arizona, Tucson, Arizona 85721, United States; Email: anubhabc@arizona.edu; orcid.org/0000-0002-1902-8308.

**Percy Zahl:** Brookhaven National Laboratory, Center for Functional Nanomaterials, Upton, New York 11973, United States; Email: pzahl@bnl.gov; orcid.org/0000-0002-6629-7500.

**Qingqing Dai:** Department of Chemistry and Biochemistry, University of Arizona, Tucson, Arizona 85721, United States; Email: qingqingdai369@gmail.com; orcid.org/0000-0003-1239-9162.

**Hong Li:** Department of Chemistry and Biochemistry, University of Arizona, Tucson, Arizona 85721, United States; Email: hongli2@arizona.edu; orcid.org/0000-0002-4513-3056.

**Torsten Fritz:** Friedrich Schiller University Jena, Institute of Solid State Physics, Helmholtzweg 5, 07743 Jena, Germany; Email: torsten.fritz@uni-jena.de; orcid.org/0000-0001-6904-1909.

**Paul Simon:** Friedrich Schiller University, Department of Solid State Physics, Helmholtzweg 5, 07743 Jena, Germany; Email: paul.simon@uni-jena.de.

**Jean-Luc Bredas:** Department of Chemistry and Biochemistry, University of Arizona, Tucson, Arizona 85721, United States; Email: jlbredas@arizona.edu; orcid.org/0000-0001-7278-4471.

## Conflicts of Interest

There are no conflicts to declare.

## Acknowledgements

This research was supported by the National Science Foundation under grant # NSF CHE-1954571 and by the College of Science of the University of Arizona. This research also used the LT-STM/HR-AFM facility of the Center for Functional Nanomaterials (CFN), which is a U.S. Department of Energy Office of Science User Facility, at Brookhaven National Laboratory under Contract No. DE-SC0012704. We also thank The University of Arizona - UA CBC Laboratory for Electron Spectroscopy and Surface Analysis (LESSA) Facility (RRID:SCR_022885) for this research.

**Supporting Information: Frontier Orbital Degeneracy: A new Concept for Tailoring the Magnetic State in Organic Semiconductor Adsorbates**

**Anubhab Chakraborty,[1] Percy Zahl,[2] Qingqing Dai,[1] Hong Li,[1] Torsten Fritz,[1,3] Paul Simon,[3] Jean-Luc Brédas [1] and Oliver L.A. Monti[1,4]*[*]**

[1]Department of Chemistry and Biochemistry, University of Arizona, Tucson, Arizona 85721, United States

[2]Brookhaven National Laboratory, Center for Functional Nanomaterials, Upton, New York 11973, United States

[3]Friedrich Schiller University Jena, Institute of Solid State Physics, Helmholtzweg 5, 07743 Jena, Germany

[4]Department of Physics, University of Arizona, Tucson, Arizona 85721, United States

S1: HATCN molecule "Bright" and "Dark" types

Constant current STM images taken at different bias and setpoint currents all show only the two types of HATCN appearances: "Bright" with a bright center, and "Dark" with a darker center (Figure S1).

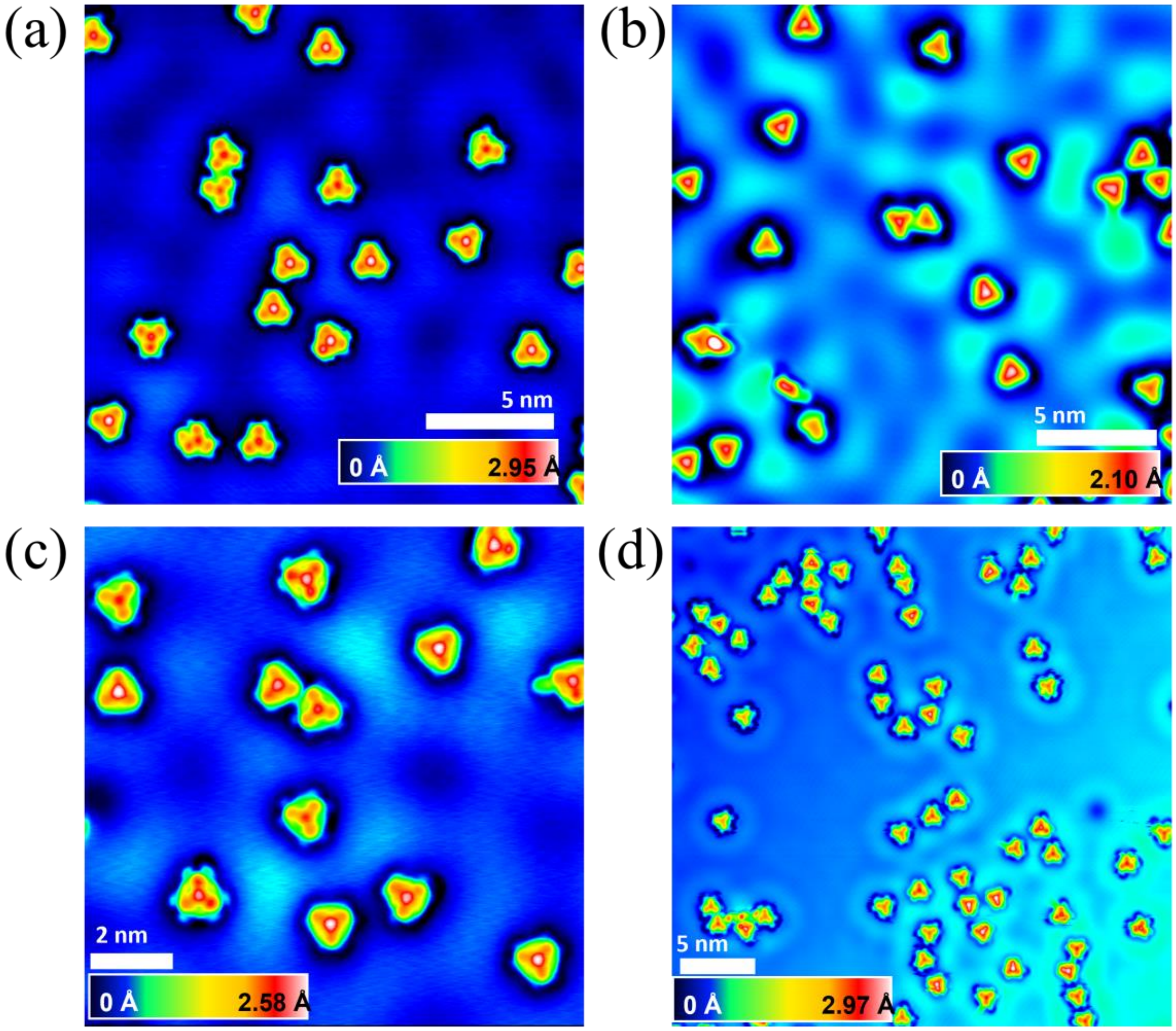

**Figure S1. Constant current STM scans, taken at scan parameters of (a) 0.05 V, 2 pA (b) 0.05 V, 20 pA (c) 0.04 V, 2 pA and (d) 0.2 V, 10 pA, each showing only the two types of HATCN appearances "Bright" and "Dark".**

S2: STM apparent height differences between "Bright" and "Dark" HATCN molecules

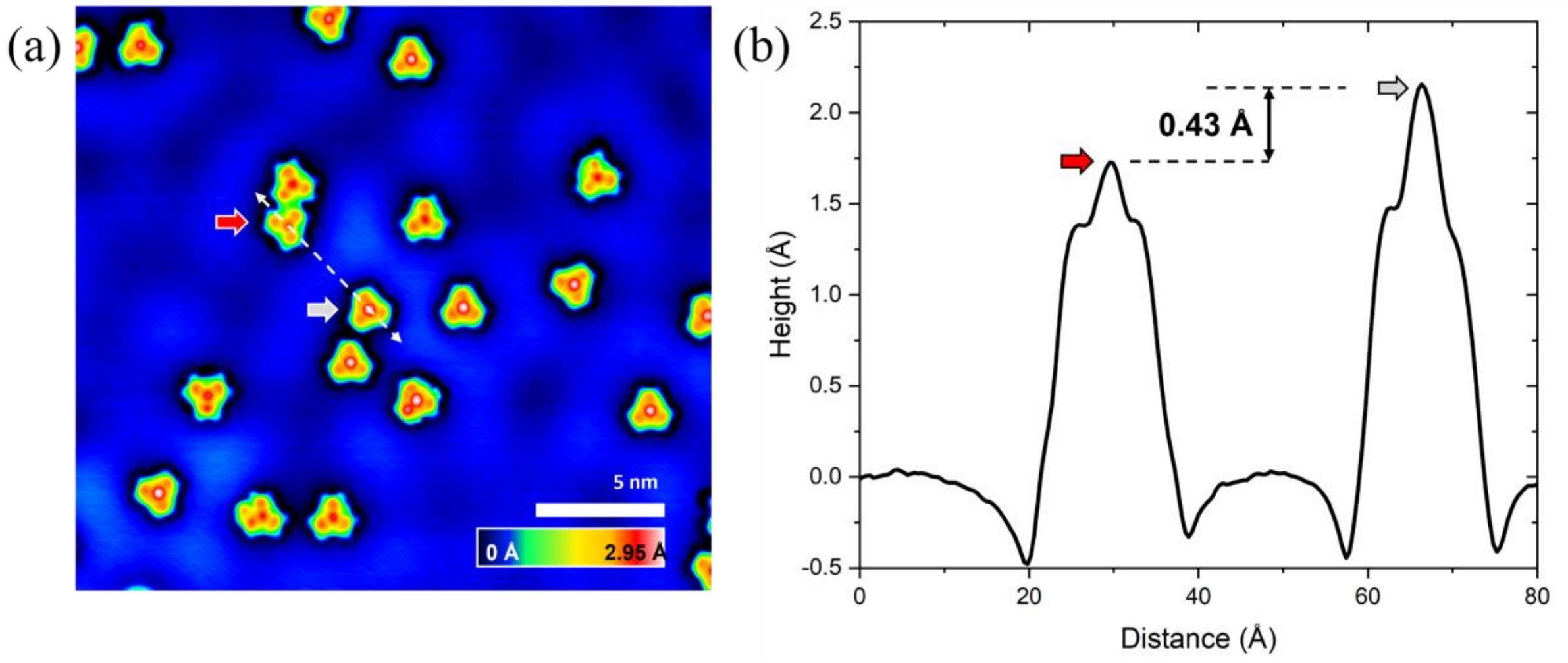

**Figure S2. (a) Constant current STM scan (0.05 V, 2 pA). (b) STM height profile taken along the dashed white line in (a), showing the STM height difference (Δh = 0.43 Å) between "Dark" (red arrow) and "Bright" (grey arrow) HATCN molecules, measured at the molecule centers.**

Figure S2 shows an example of the STM apparent height difference between "Bright" and "Dark" molecules, with the heights measured at the center of the molecule. Background correction of the STM images was done in GXSM to account for the *z*-drift of the STM tip over time during the scanning process.

## S3: The zero-bias feature is an ASK resonance

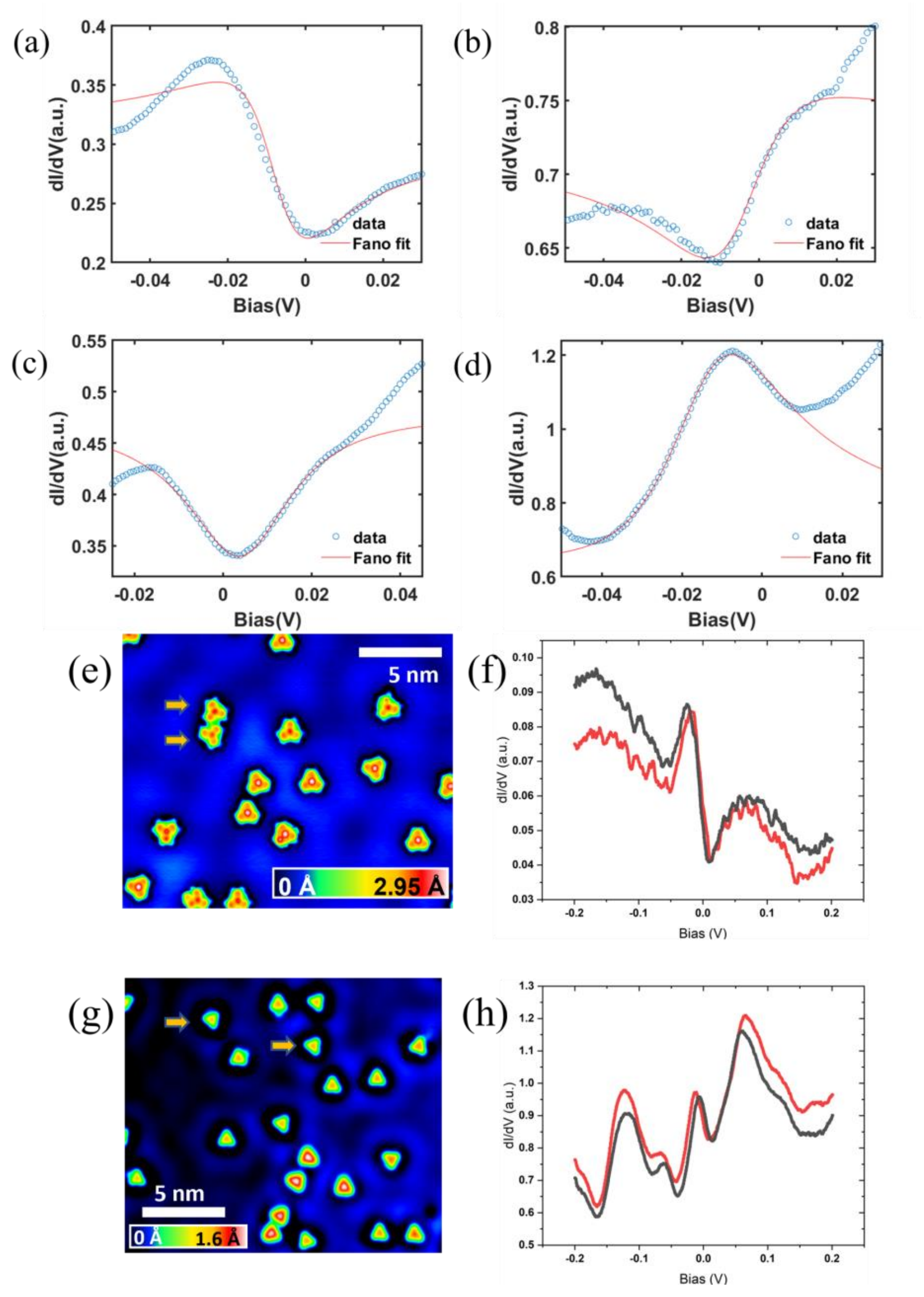

**Figure S3. Fano fits for the ASK resonance features observed in the $\frac{dI}{dV}$ spectra, measured over different "Dark" HATCN molecules, with the Fano coupling parameter (*q*) values of (a) -0.7, (b) 0.6, (c) 0.06 and (d) -0.3. Constant current STM scans at (e) (0.05 V, 2 pA) and (g) (0.2 V, 0.01 nA), and corresponding $\frac{dI}{dV}$ spectra (f) and (h) of "Dark" molecules highlighted with arrows.**

The zero-bias feature in STS exhibits different peak shapes for different “Dark” HATCN molecules (Fig S3a-S3d). We capture this striking behavior by varying the Fano coupling parameter $q$, found by fitting the $\frac{dI}{dV}$ spectra of 35 “Dark” HATCN molecules with the Fano function (equation (1) of the main text). $q$ falls between ±1 in our measurements. The origin of this resonance shape variation is most likely due to different tip conditions between scans over different regions on the surface, which is reflected in STM imaging as well, where the appearance of the HATCN molecules shows different levels of detail (cf. Fig S3e vs. S3g). Moreover, even though the ASK resonance shape varies between measurements taken over different regions on the surface due to varying tip conditions, the shapes (and $q$) are nearly identical for $\frac{dI}{dV}$ measurements on molecules in the same region (Fig S3f and S3h). The observation of different shapes for the zero-bias feature in different STS spectra also suggests that this is indeed an ASK resonance in the coupled HATCN/Ag(111) system and not a spectral feature of either HATCN or Ag(111), and that slight changes to the tip and scan conditions determine the precise appearance of the ASK resonance.

S4: On-surface orientation of HATCN molecules

Figure S4a shows different on-surface rotational orientations of HATCN on Ag(111). This and other such images show that the “Bright” and “Dark” types are not correlated with the molecular orientation on the Ag(111) surface. Figure S4b and S4c show that the switching process is not caused by the molecules moving to a different adsorption site, which would require a displacement on the order of at least 1 Å.

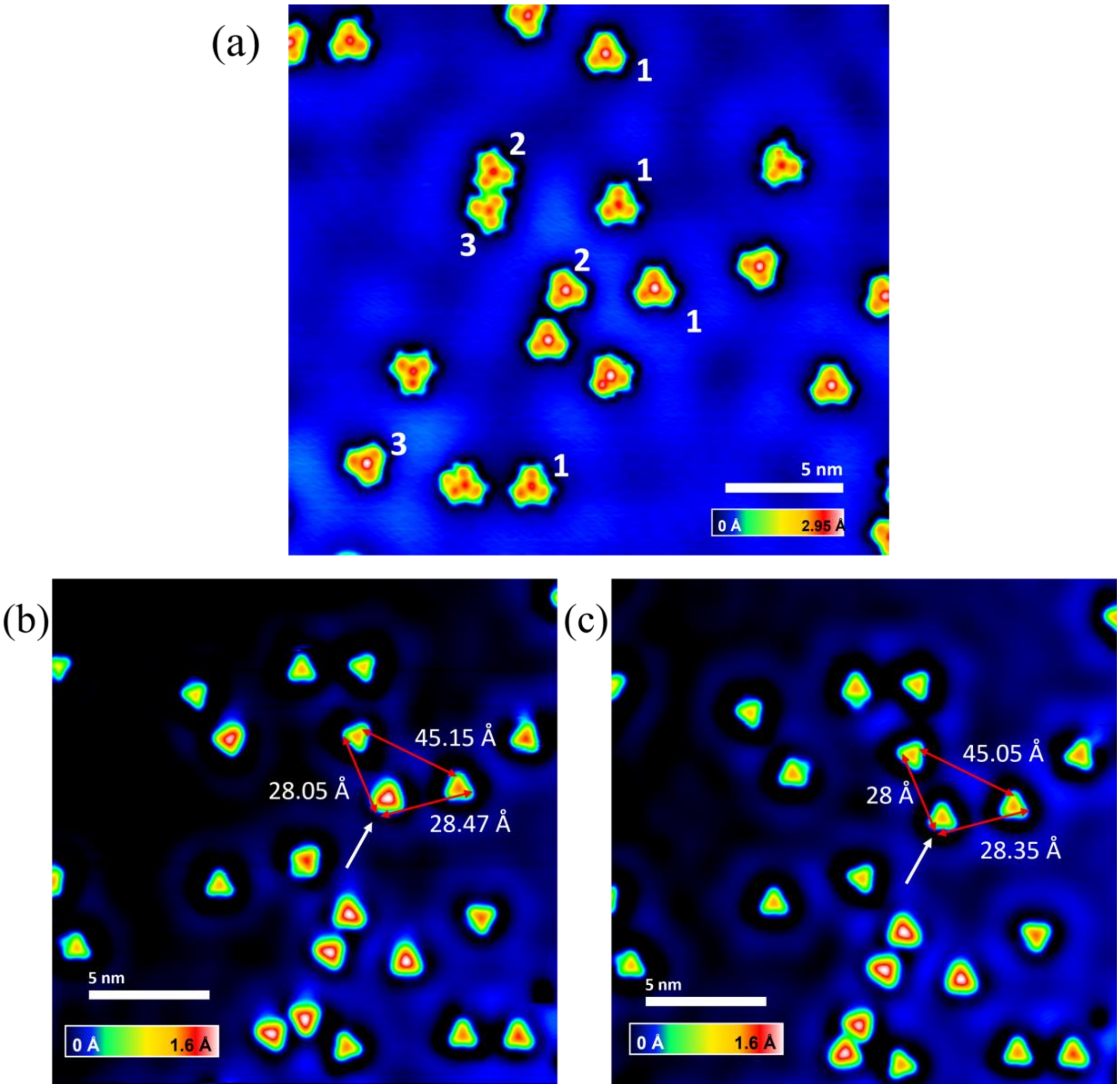

**Figure S4. (a) Constant current STM scan (0.05 V, 2 pA), showing the two molecular “types” and relative orientations of the HATCN molecules on Ag(111). Three observed orientations are labeled 1-3 alongside molecules of different “types”. Note that the labelling scheme is only relative to other molecules in this particular STM image, and not absolute since the underlying Ag (111) lattice orientation is not known. Constant current STM scans (0.2 V, 0.01 nA) (b) before and (c) after the molecule shown with white arrow switches from “Bright” to “Dark”. Center-to-center distances of this molecule to two other molecules shows that its position has not changed significantly during the switching process.**

## S5: Low Energy Electron Diffraction (LEED) of 1 ML HATCN/Ag(111)

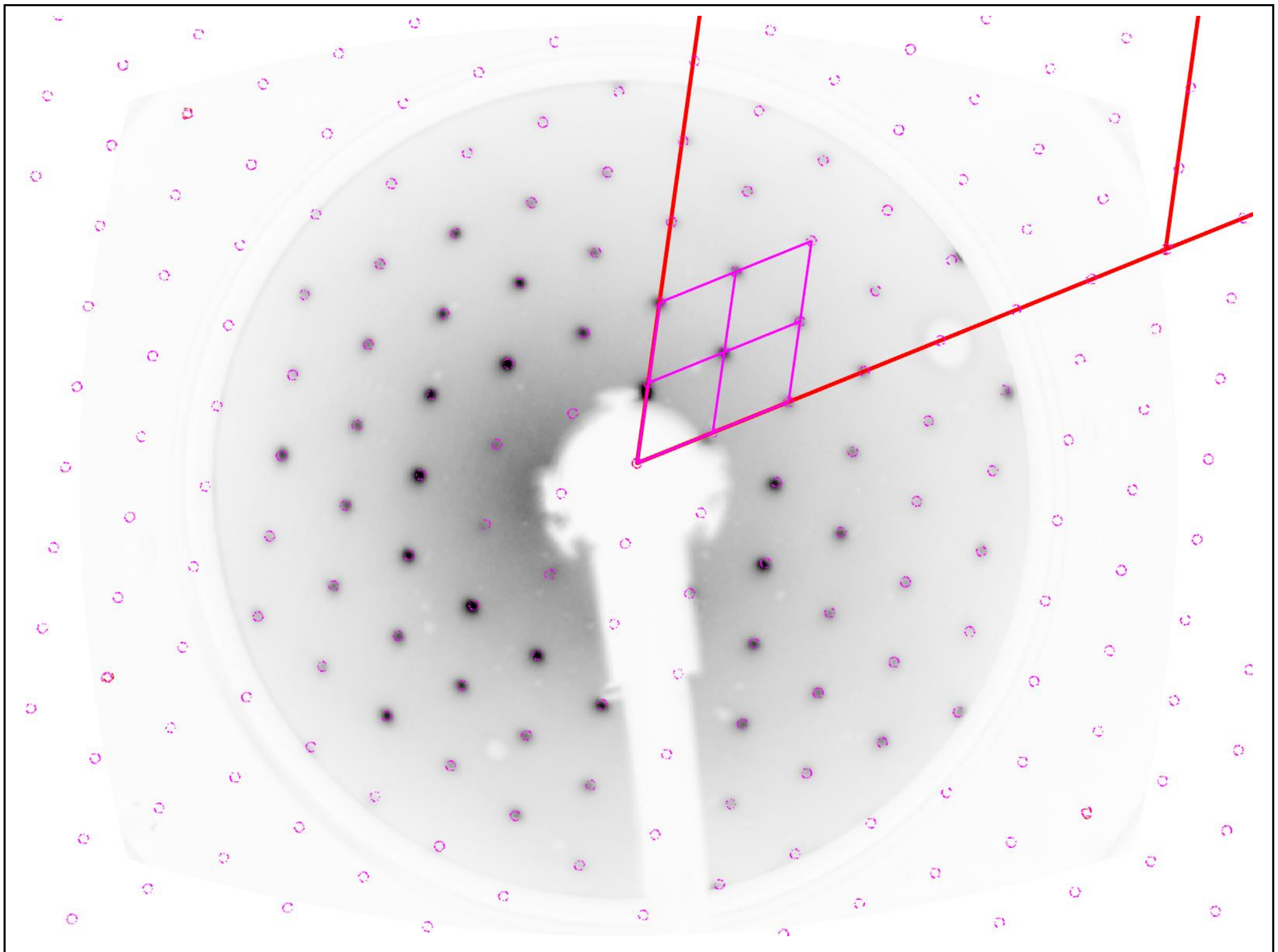

**Figure S5. Contrast-inverted, distortion-corrected LEED image of 1 ML HATCN/Ag(111). The image was acquired at 41.5 eV at room temperature. Lines depict the 2D reciprocal unit cell of the Ag(111) substrate (red) and that of the HATCN 1 ML film (magenta), showing the (7x7) superstructure of the film.**

In addition to LT-STM, Low Energy Electron Diffraction (LEED) measurements were performed to investigate the epitaxial film structure for 1 ML HATCN/Ag(111). Figure S5 shows the distortion-corrected[1] LEED image of 1 ML HATCN/Ag(111), showing the formation of a commensurate (7x7) superstructure for the HATCN film, as already reported previously.[2–4]

S6: STS measurements on 1 ML HATCN/Ag(111)

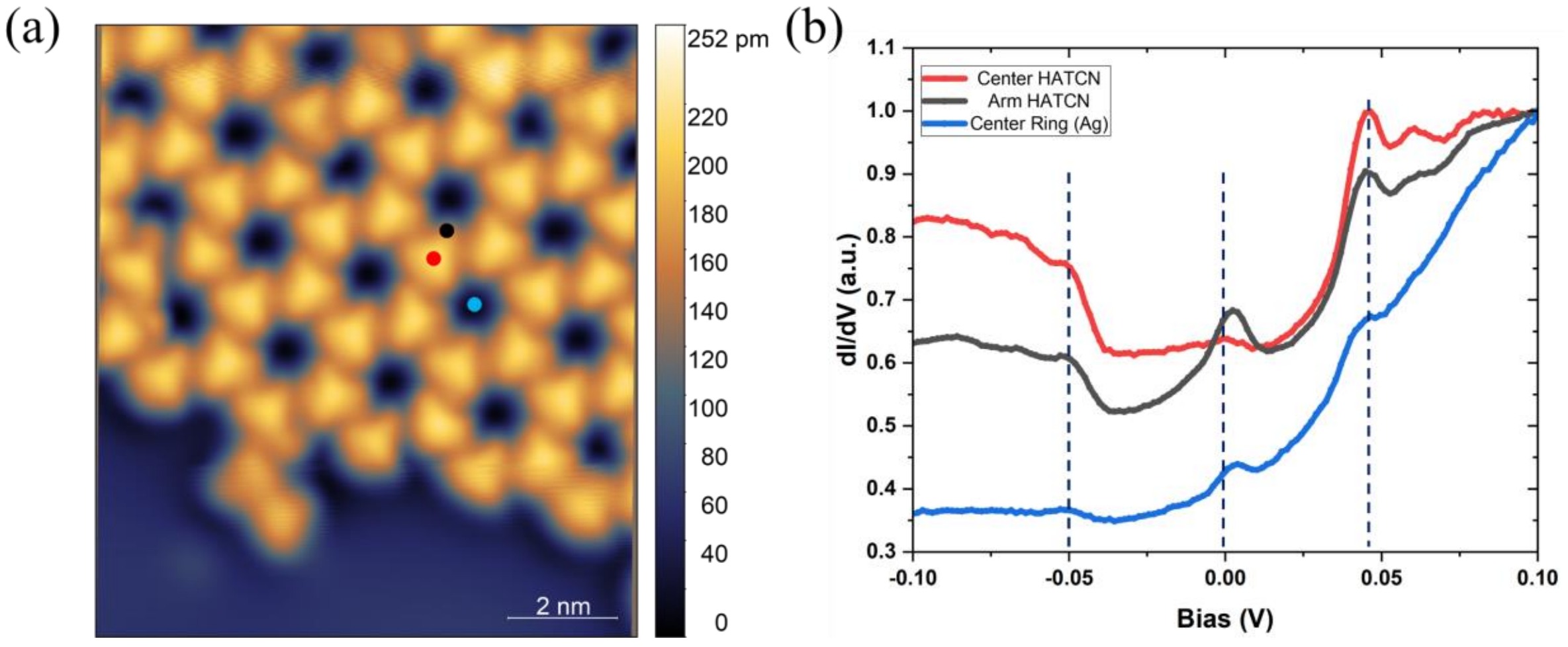

**Figure S6. (a) Constant current STM image (-0.4 V, 0.7 nA, 4.5 K) of 1 ML HATCN/Ag(111). The colored dots indicate the positions where the $\frac{dI}{dV}$ spectra shown in (b) were measured. (b) $\frac{dI}{dV}$ spectra taken over different positions, with the dashed lines indicating the features of importance at -50 mV, 0 mV and +46 mV.**

Figure S6a shows the molecular self-assembly of HATCN on Ag(111). Figure S6b shows $dI/dV$ spectra measured over 3 different positions on 1 ML HATCN/Ag(111). The small peak at -50 mV is attributed to the Ag(111) Shockley surface state. The small peak near 0 mV is most likely from the partially occupied former LUMO of HATCN (CT Singlet $|\alpha_1\rangle$ (or $|\beta_1\rangle$)), while the stronger feature at around +46 mV is likely from the fully unoccupied LUMO of HATCN (similar to CT Singlet $|\alpha_2\rangle$ (or $|\beta_2\rangle$)). Note that these features are also observed weakly in the STS spectra taken over Ag(111) in the pores of the HATCN lattice (blue curve, figure S6b), most likely because of a residual contribution to the Local Density of States from the 6 surrounding HATCN molecules. The observation of the feature near 0 mV at different positions over the HATCN molecule in the close packed layer and even over Ag(111), as well as its weaker intensity (and no variation in shape) compared to the ASK resonance features leads us to conclude that the molecules in close-packed layers are similar to the isolated “Bright” molecules and that the near 0 mV features observed in both these cases are due to the existence of the CT Singlet $|\alpha_2\rangle$ (or $|\beta_2\rangle$) state near $E_F$.

S7: Electronic structure of an isolated HATCN molecule

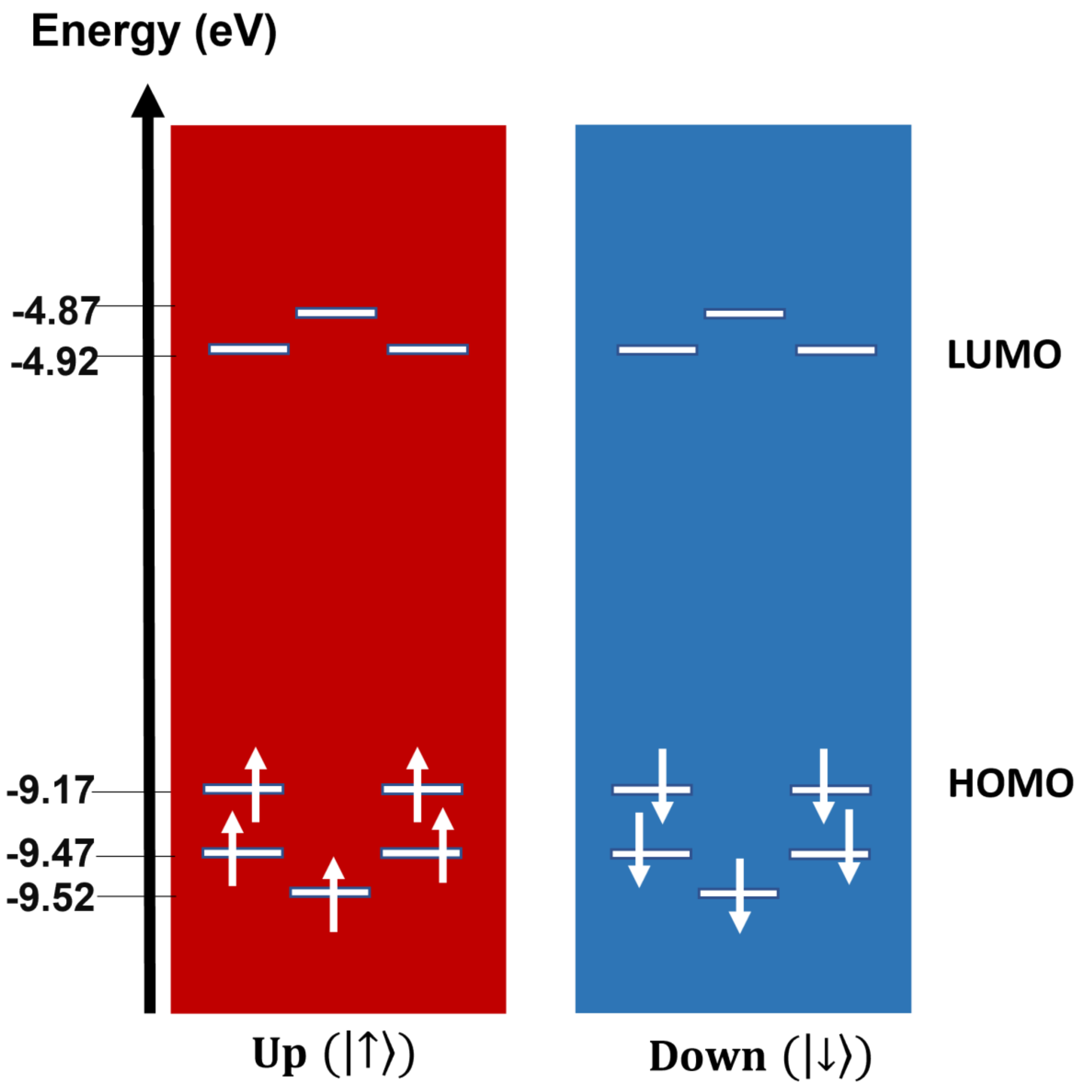

Figure S7. Calculated energy level diagram of the frontier molecular orbitals for an isolated HATCN molecule, shown for each spin channel.

Figure S7 shows the DFT energy levels for the frontier molecular orbitals of an isolated neutral HATCN molecule, calculated at the B3LYP / 6-311++G level. Both HOMO and LUMO are doubly degenerate.

S8: DFT calculations on the HATCN/Ag(111) interface with consideration of different adsorption sites

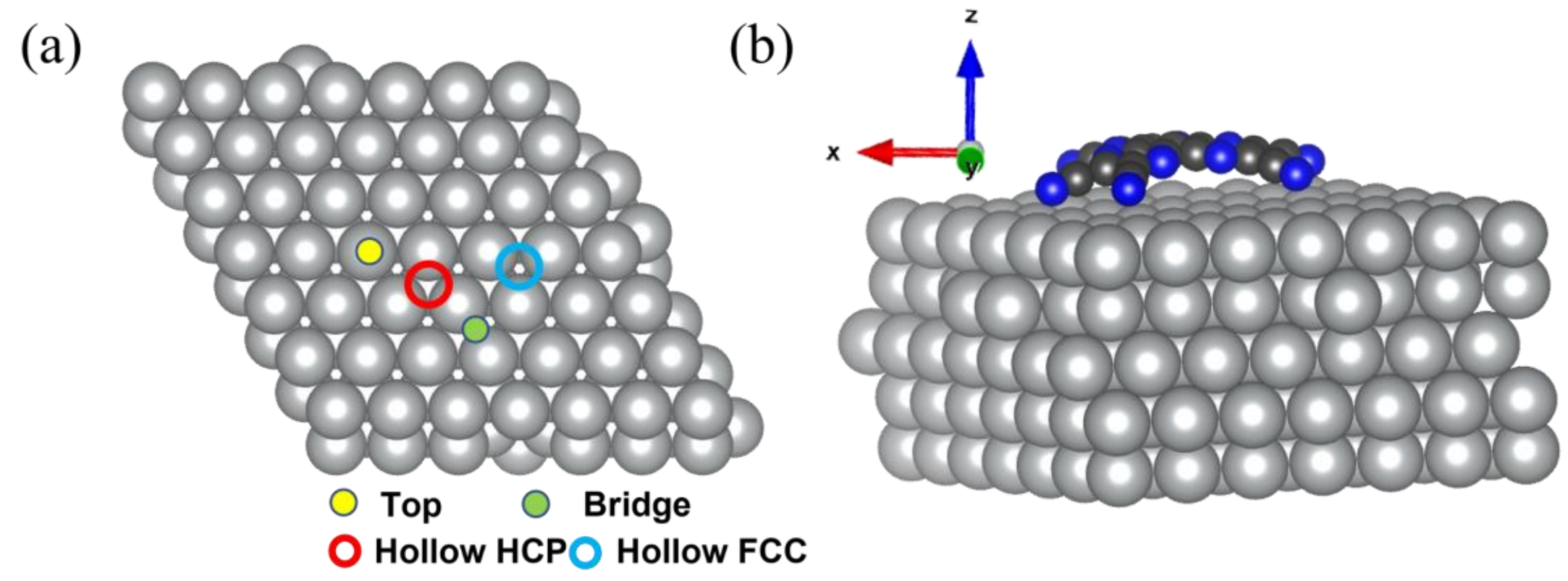

**Figure S8. Illustration of (a) the sites on Ag(111) considered in our DFT calculations and (b) the buckling of the HATCN molecule on Ag(111) caused by bending of the -CN groups towards the surface.**

**Table S1. Calculated adsorption site dependent properties for HATCN/Ag(111)**

| Adsorption Site | Adsorption Energy (eV/unit cell) | Adsorption Height (Å) | Magnetic Moment/$\mu_B$ |
|---|---|---|---|
| **Top** | -2.97 | 2.6 | 0.49 |
| **Hollow hcp** | -3.04 | 2.6 | 0.61 |
| **Hollow fcc** | -3.05 | 2.6 | 0.62 |
| **Bridge** | -2.97 | 2.6 | 0.56 |

Table S8 lists the adsorption energy, adsorption height, and magnetic moments for an HATCN molecule adsorbed on four different sites of the Ag(111) surface, calculated using the PBE functional with DFT-D3 dispersion corrections (see DFT Methodology). The adsorption energies are very similar, and the adsorption heights are identical for all 4 sites considered. Note that the adsorption height is calculated as the vertical distance between the C atoms in the central benzene ring of HATCN and the Ag atoms of the topmost Ag layer. The most stable adsorption site is the hollow fcc site. The fact that the various adsorption sites have very similar adsorption energies, adsorption heights, and magnetic moments suggests that adsorption at different sites is not responsible for the appearance of the two molecular contrasts in STM. This conclusion is also supported by the fact that there is no correlation between molecular orientation on the surface and whether a molecule appears as “Bright” or “Dark”.

S9: Projected density of states (PDOS) calculated at the DFT-PBE level for different adsorption sites of HATCN on Ag(111)

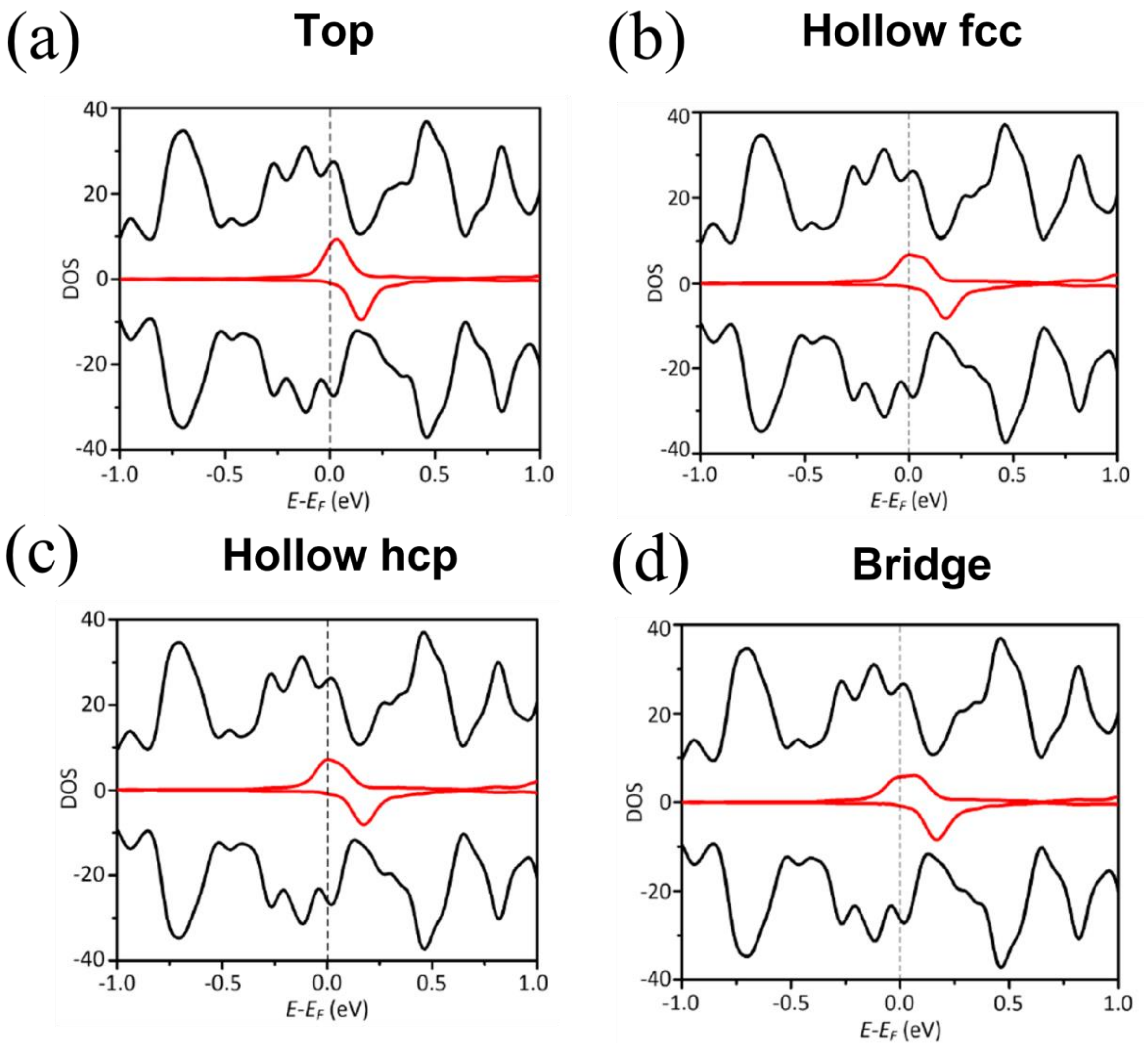

**Figure S9. Calculated PDOS of HATCN/Ag(111), shown for both spin channels for (a) top, (b) hollow fcc, (c) hollow hcp, and (d) bridge sites. Black curves: total DOS; red curves: HATCN PDOS.**

The HATCN PDOS curves shown in Figure S9 (red curves) demonstrate that the electronic structure near $E_F$ is similar for all 4 adsorption sites and that they all share the general features of the CT Doublet state shown in Figure 5a. SDFT calculations were specifically performed for the Hollow FCC and the Bridge sites, and both sites show the existence of the CT Doublet and CT Singlet states with similar energetic difference. These results imply that the existence of the CT Doublet and CT Singlet states is not a result of adsorption on different sites of Ag(111).

<u>S10: Gaussian fits of the HATCN PDOS</u>

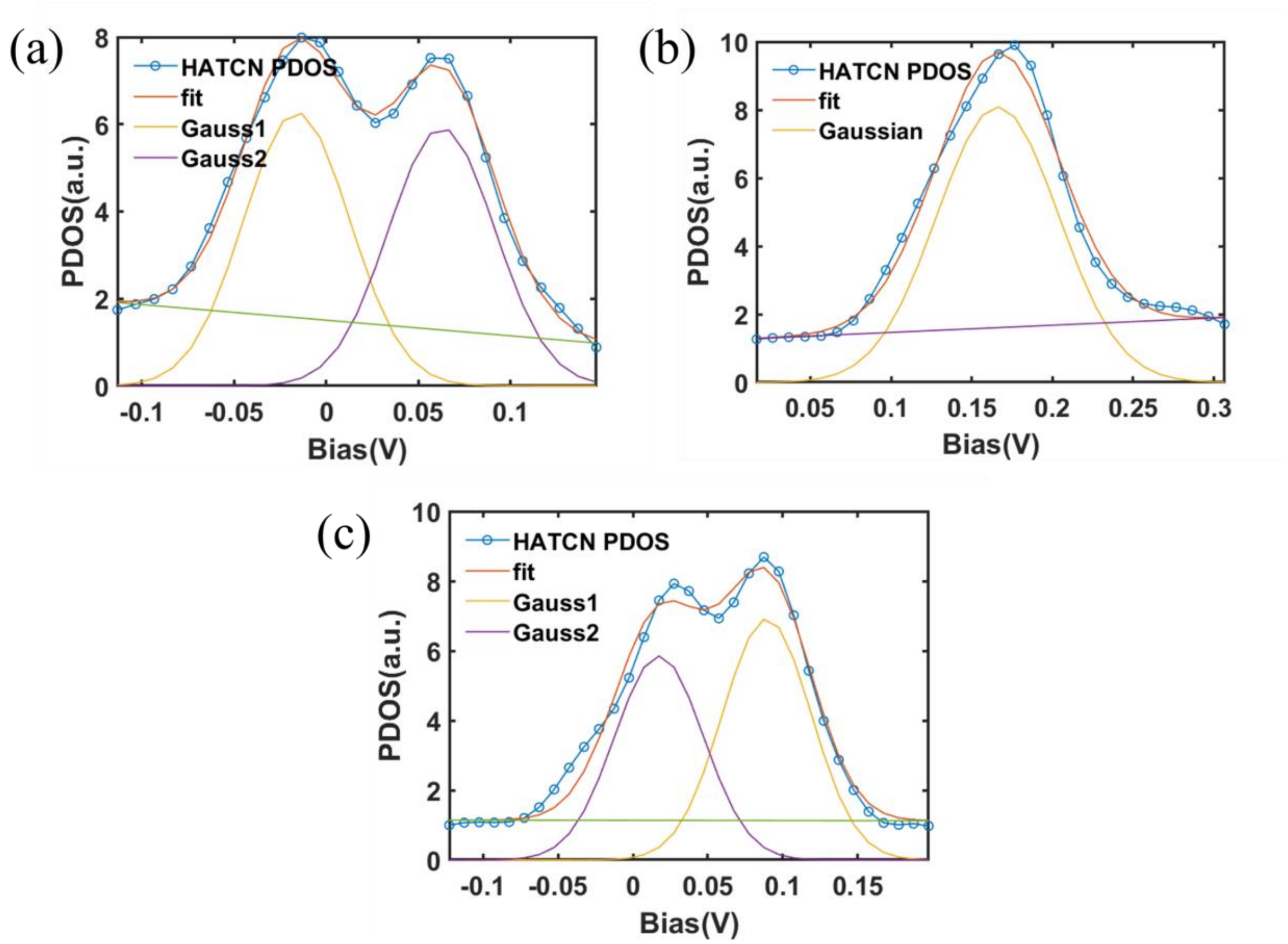

**Figure S10. Gaussian fits of the calculated PDOS of HATCN (red curves in Figure 5a, b) near $E_F$, shown for the (a) CT Doublet Up ($|\uparrow\rangle$), (b) CT Doublet Down ($|\downarrow\rangle$), and (c) CT Singlet Up ($|\uparrow\rangle$) channels. Gauss1 and Gauss2 refer to the 2 Gaussian functions used for the fits (equation *1*); green lines in (a) and (c) and the purple line in (b) are the linear background contribution to the fits. The CT Singlet Down ($|\downarrow\rangle$) state is degenerate and identical to the CT Singlet Up ($|\uparrow\rangle$) state, and hence the fit for this channel is not shown here separately.**

The HATCN PDOS near $E_F$ for each spin channel for the CT Doublet and CT Singlet states were fitted using the following expression:

$$f(x) = a_1 * \exp\left(-\frac{x-b_1}{c_1}\right)^2 + a_1 * \exp\left(-\frac{x-b_2}{c_1}\right)^2 + D * x + E \qquad (1)$$

The two Gaussians in the equation account for the 2 former LUMOs of isolated HATCN, one of which is singly occupied upon a 1-electron transfer from Ag(111). In addition, we include a linear background term to capture residual contributions to the PDOS in the small fitting window. Note that for the CT Doublet Down state, only a single Gaussian function was used for the fit, since the two former LUMOs remain degenerate (perfect correlation coefficient of 1 for their energies $b_1$ and $b_2$ when fitted using equation *1*). Table S2 lists the fitted parameters for each state.

Table S2. Fitted parameters from HATCN PDOS Gaussian fits

| State | $b_1$ (meV) | $b_2$ (meV) | $c_1$ (meV) |
|---|---|---|---|
| CT Doublet Up | -16(2) | 63(2) | 40(4) |
| CT Doublet Down | 166(1) | 166(1) | 53(2) |
| CT Singlet Up | 17(4) | 89(4) | 42(7) |
| CT Singlet Down | 17(4) | 89(4) | 42(7) |

The values from Table S2 were considered to draw the energy-level diagram in Figure 6 in the main text. The area under the fitted curve below $E_F$ was used to obtain the electronic charges for each of the 4 states, via:

$$Electronic\ charge = \frac{Area\ below\ E_F}{Total\ area} * 1\ e \tag{2}$$

The calculated electronic charges for each state are reported in Figure 6 of the main text, showing the charge distribution for the CT Doublet and CT Singlet states.

## S11: HR-AFM contrasts and comparison with STM

Figure S11 shows that there is no correlation between molecules with observed contrast differences in STM and the corresponding HR-AFM scans, and in fact, there is no consistent contrast difference between different HATCN molecules of the same type in the HR-AFM images themselves.

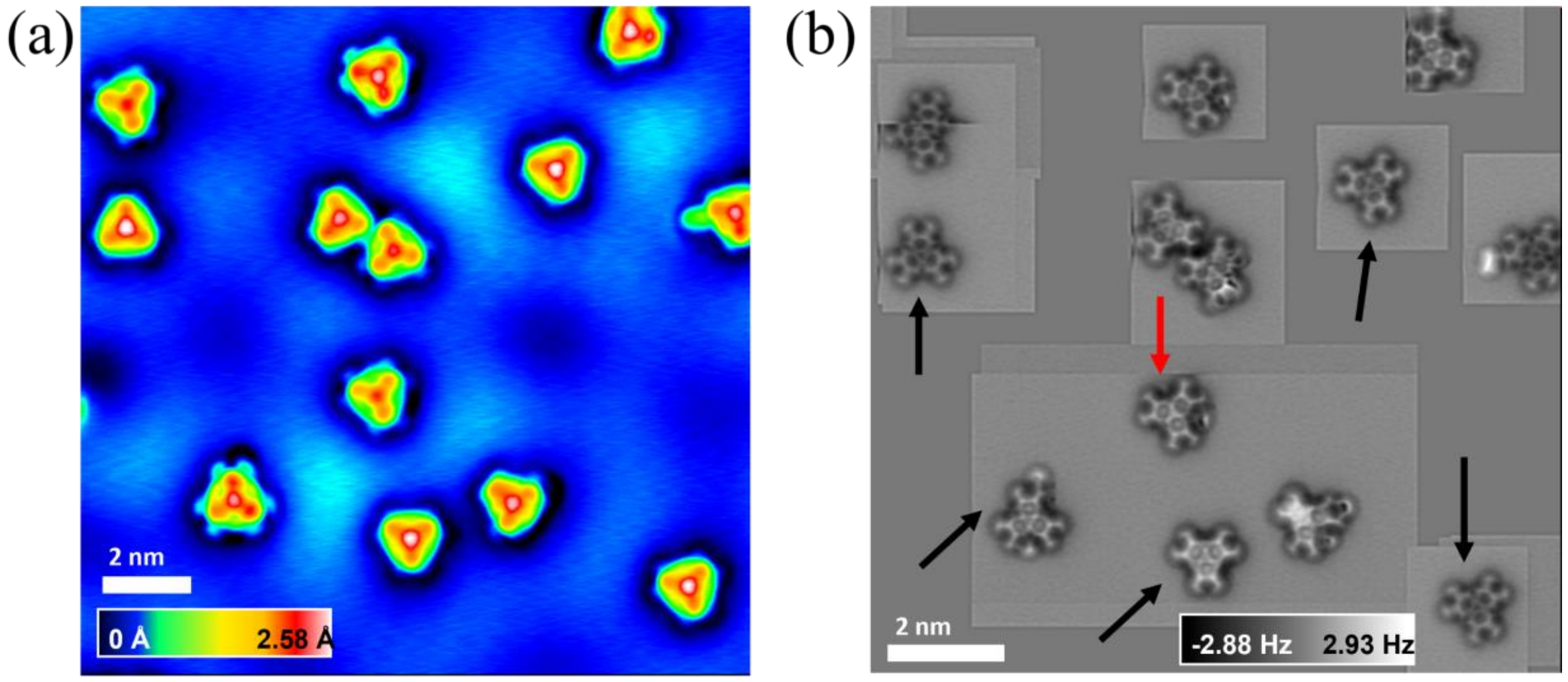

**Figure S11. (a) Constant current STM (0.04 V, 2 pA) and (b) HR-AFM (0.02 V) of the same region. The black and red arrows in (b) indicate "Bright" and "Dark" molecules, respectively. It is readily seen that different "Bright" molecules have different contrasts in HR-AFM scans, and some "Dark" and "Bright" molecules have very similar contrasts as well.**

S12: STM simulations of CT Doublet and CT Singlet states

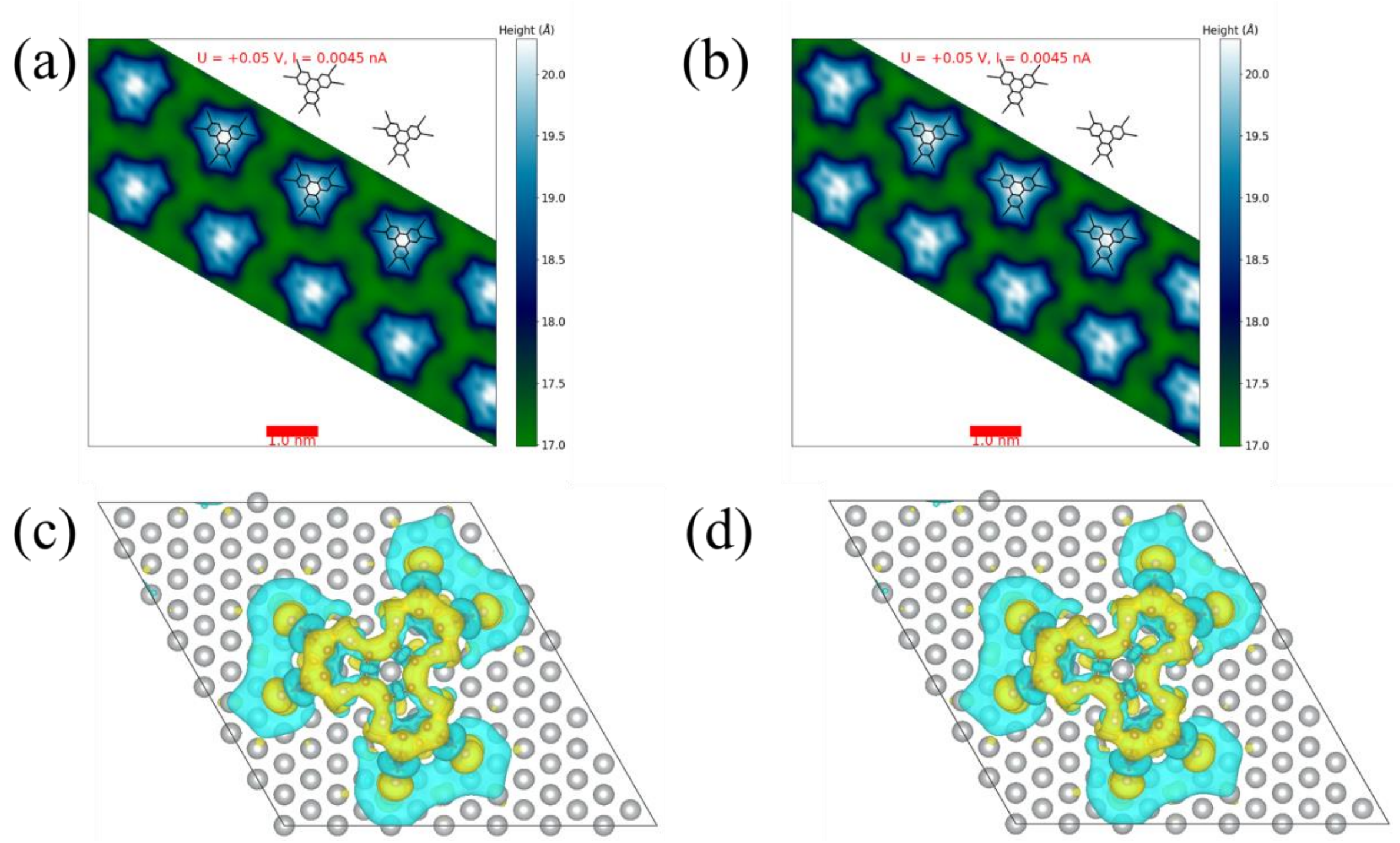

**Figure S12. STM simulations (0.05 V, 4.5 pA) of (a) CT Doublet and (b) CT Singlet states of HATCN/Ag(111). Charge density distribution plots of (c) CT Doublet and (d) CT Singlet states, plotted at the same isosurface value.**

STM simulations (Figure S12a and S12b) of the CT Doublet and CT Singlet states of HATCN/Ag(111) do not show the contrast differences seen in the constant current experimental STM scans due to almost identical calculated charge density distributions (Figure S12c and S12d) of the two states.